\documentclass[a4paper,fleqn,usenatbib]{mnras}

\usepackage{graphicx}	
\usepackage{amsmath}	
\usepackage{amssymb}	
\usepackage{rotating}
\usepackage{pdflscape}
\usepackage{longtable}
\usepackage{afterpage}
\usepackage[T1]{fontenc}

\title[Identification of ${\nu}_6$ asteroids]
      {Identifying the population of stable ${\nu}_6$ resonant asteroids using large databases}
       \author[V. Carruba, S. Aljbaae, R. C. Domingos, M. Huaman, B. Martins]{V. Carruba$^{1}$\thanks{E-mail: valerio.carruba@unesp.br}, S. Aljbaae$^{2}$, R. C. Domingos$^{3}$, M. Huaman$^{4}$, B. Martins$^{1}$.\\
         $^{1}$S\~{a}o Paulo State University (UNESP), School of Natural Sciences and Engineering, Guaratinguet\'{a}, SP, 12516-410, Brazil \\
         $^{2}$National Space Research Institute (INPE), Division of Space Mechanics and Control, C.P. 515, 12227-310, S\~{a}o Jos\'e dos Campos, SP, Brazil \\
         $^{3}$S\~{a}o Paulo State University (UNESP), S\~{a}o Jo\~{a}o da Boa Vista, SP, 13876-750, Brazil \\
         $^{4}$Universidad tecnol\'{o}gica del Per\'{u} (UTP), Cercado de Lima, 15046, Per\'{u} \\
}

       \date{Accepted 2022 June 14. Received 2022 June 14; in original form 2022 March 30}
       \pubyear{2019}

\begin{document}
\label{firstpage}
\pagerange{\pageref{firstpage}--\pageref{lastpage}} 
\maketitle

\begin{abstract}

  Large observational surveys, like those that will be conducted at the Vera
  C. Rubin Observatory, are expected to discover up to one million new
  asteroids in the first year of operation. This will more than double the
  database of known asteroids. New methods will be needed to handle the
  large influx of data. Here, we tested some of these new approaches by
  studying the population of asteroids on stable orbits inside the ${\nu}_6$
  secular resonance. This resonance is one of the strongest mechanisms for
  destabilizing the orbits of main-belt bodies and producing Near-Earth
  Asteroids (NEAs). Yet, stable orbital configurations where the asteroid
  pericenter is either aligned or anti-aligned with that of Saturn exist
  inside the resonance. The population of stable ${\nu}_6$ resonators is now
  the largest population of asteroids in stable orbits inside a secular
  resonance. Here we obtained the largest sample of asteroids’ proper elements
  ever used.   Clustering methods and the use of machine learning algorithms
  permitted the identification of the known asteroid families crossed by
  the ${\nu}_6$ resonance  and of two entirely new groups: the Tiffanykapler
  and the 138605 QW177 families. The Tiffanykapler family is the first
  young asteroid family ever found in a linear secular resonance, with an
  age of $3.0\pm1.2$ My and an ejection velocity field parameter of
  $V_{EJ} = 15^{+6}_{-3}$ m/s. We identify a population of high-eccentricity
  objects around the Tina family that may be the first example of an
  asteroid family ``resonant halo''.
\end{abstract}


\section{Introduction}
\label{sec: intro}

Linear secular resonances occur when we have a commensurability between the
frequency of precession of an asteroid perihelion, $g$, or node, $s$, and
that of a planet.  The ${\nu}_6$ occurs when there is a commensurability
between the frequency $g$ of an asteroid and the $g_6$ frequency of Saturn.
This resonance is important for the effect it has on the orbital distribution
of asteroids: since it is a pericenter resonance, it can increase most
asteroids' eccentricities to planet-crossing
levels while keeping their inclinations approximately constant. It can,
therefore, destabilize most of the bodies that interact with it, and it
is a major source of NEAs through Yarkovsky effect
\citep{2000Sci...288.2190B}. Finally, in the
central main-belt, between the 3-1 and 5-2 mean-motion resonances with
Jupiter, {\bf it} separates the population of highly inclined asteroids from
the more numerous low-inclination ones \citep{2010MNRAS.408..580C}.

\begin{figure*}
  \centering
    \begin{minipage}[c]{0.45\textwidth}
    \centering \includegraphics[width=2.3in]{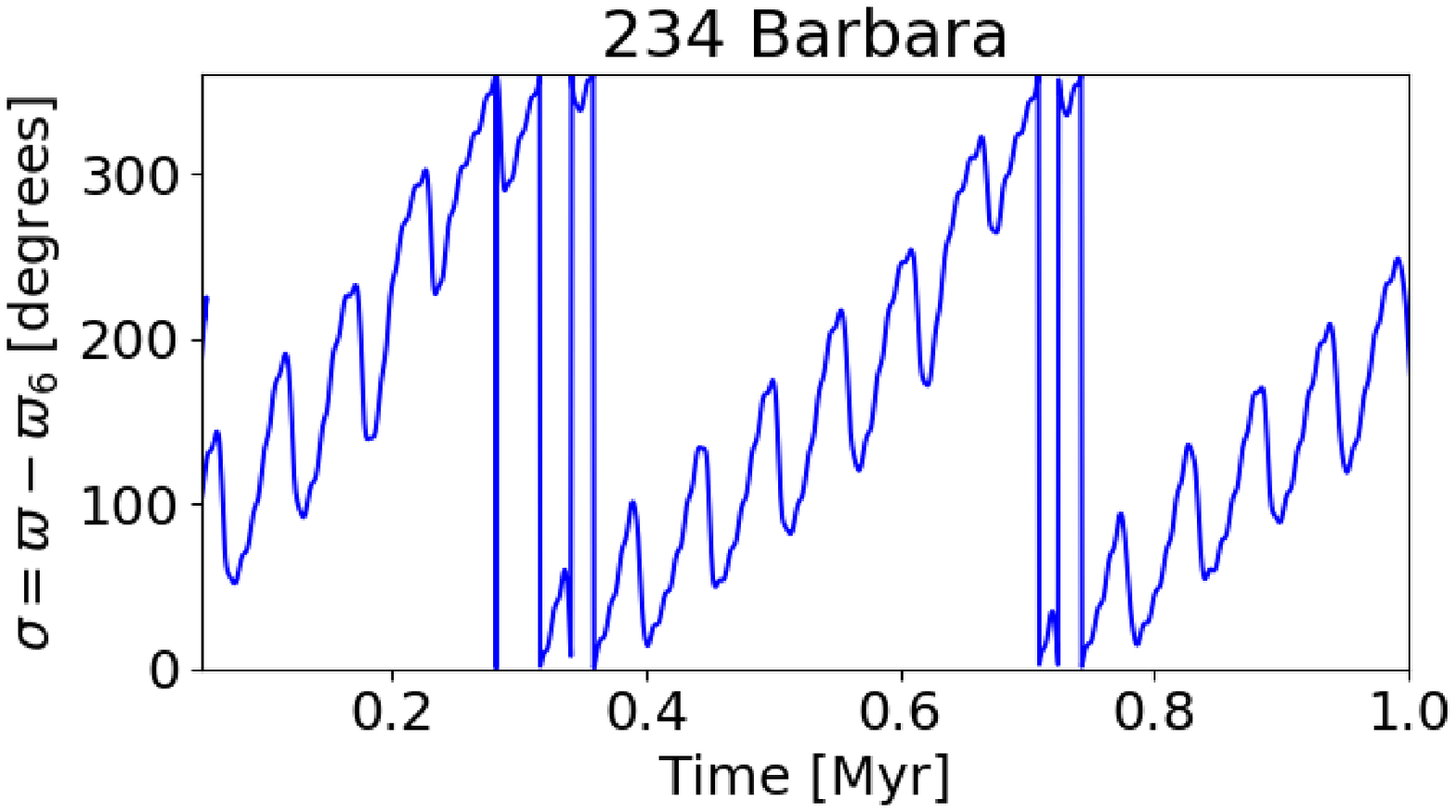}
  \end{minipage}%
  \begin{minipage}[c]{0.45\textwidth}
    \centering \includegraphics[width=2.3in]{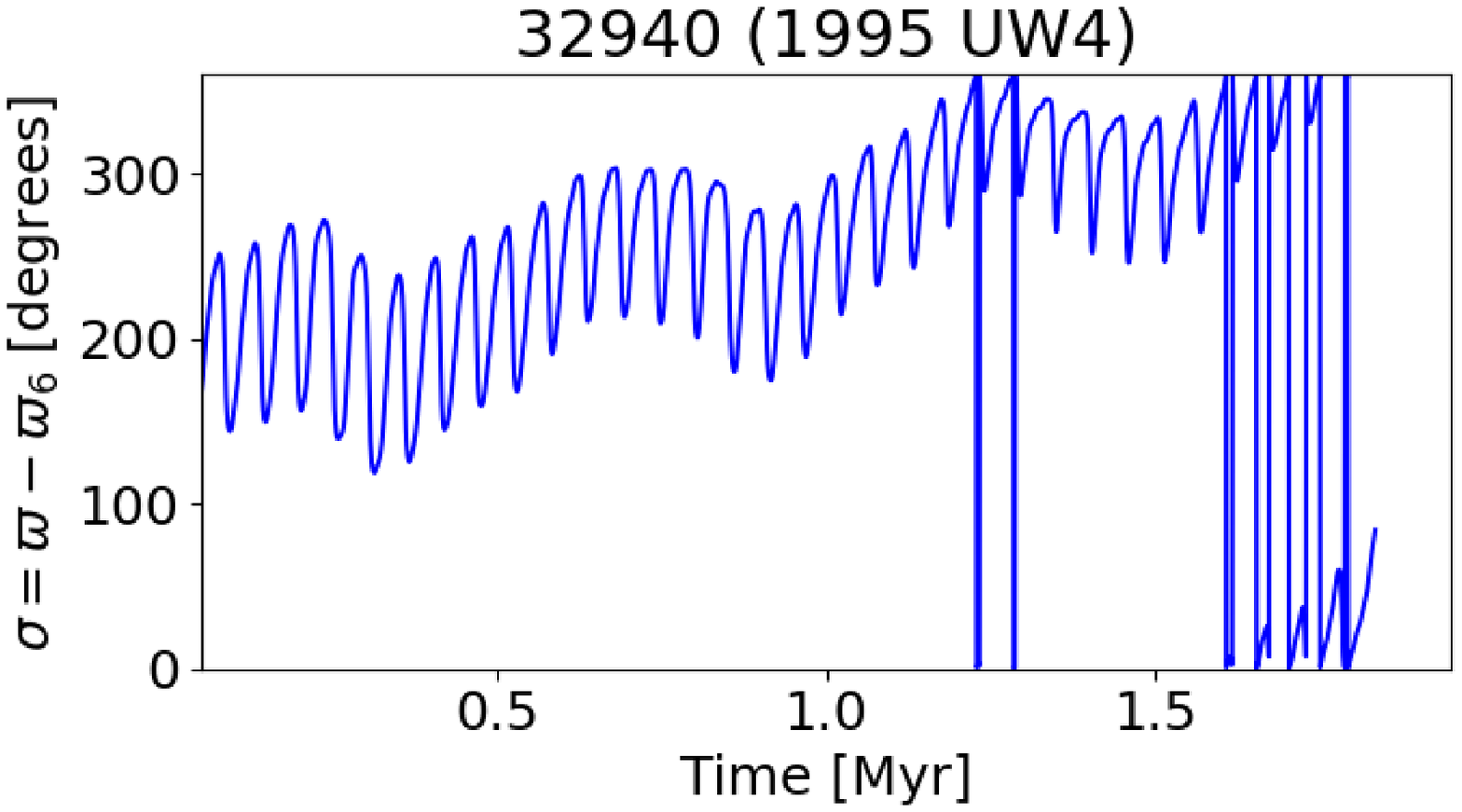}
  \end{minipage}
  \begin{minipage}[c]{0.45\textwidth}
    \centering \includegraphics[width=2.3in]{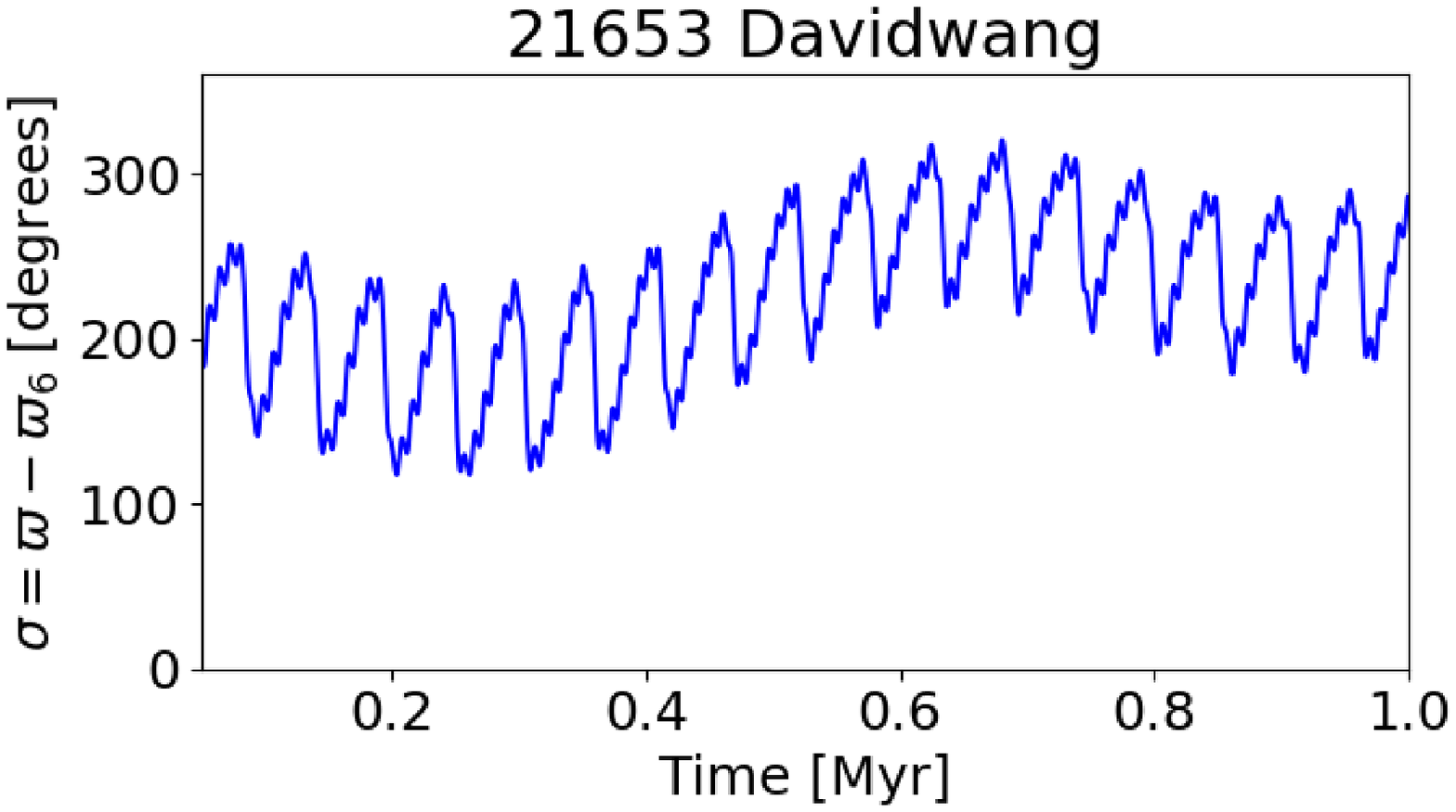}
  \end{minipage}%
    \begin{minipage}[c]{0.45\textwidth}
    \centering \includegraphics[width=2.3in]{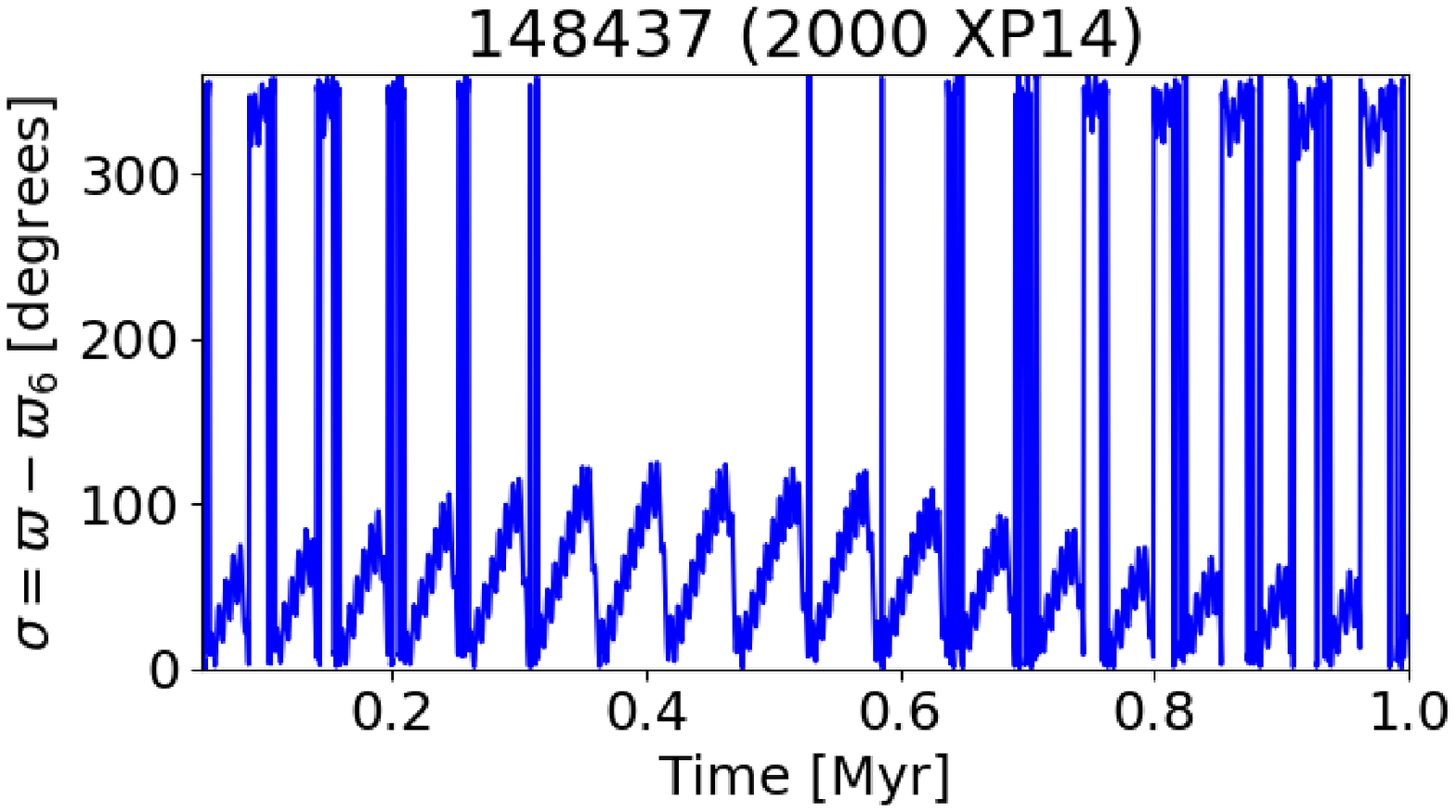}
  \end{minipage}
  \caption{Time behavior of the resonant angle $\varpi -{\varpi}_6$ for
    an asteroid in a circulating orbit (top left panel), in a switching orbit
    (top right panel) in an anti-aligned orbit (bottom left panel) and in an
    aligned orbit (bottom right panel).}
\label{fig: res_angle_ex}
\end{figure*}

While most of the asteroids interacting with the ${\nu}_6$ secular resonance
are dynamically unstable on timescales of a few My or less, islands
of dynamical stability can be found inside the resonance itself.
\citep{1987CeMec..40..233Y}, \citep{1991Icar...93..316K},
\citep{1991CeMDA..51..169M} and \citep{1993Icar..105...48M}
used a Hamiltonian technique to study the dynamics and topology of
the phase space of linear secular resonances in detail. 
The dynamics of the ${\nu}_6$ secular resonance is characterized
by the libration or circulation of the resonant angle $\sigma =
{\varpi}-{\varpi}_6$, with ${\varpi}$ the longitude of pericenter
of the asteroid and ${\varpi}_6$ that of Saturn.  If the asteroid
is outside the resonance, the resonant angle will vary continuously
between $0^{\circ}$ and $360^{\circ}$.  In a ${\nu}_6$ librating state
the resonant angle will oscillate around an equilibrium point,
which can either be $0^{\circ}$ or $180^{\circ}$.  In the first case,
we have an ``aligned libration'', since the pericenters of both
Saturn and the asteroid are pointing in the same direction.
For the second case, we have an ``anti-aligned libration'', and the
two pericenters point in opposite directions. Examples of the time
behavior of the resonant angle of real asteroids in circulating,
switching, anti-aligned, and aligned libration states are shown in
figure~(\ref{fig: res_angle_ex}).

Anti-aligned libration
is more common near the planet's orbit, and, indeed, is more common
in the central and outer main-belt. \citet{2011MNRAS.412.2040C} identified
the first case of an asteroid family completely made of asteroids
in anti-aligned libration states, that of 1222 Tina.
\citep{2014ApJ...792...46C}
later identified the case of the Euphrosyne family, with the second-largest
population of asteroids in such states. Finally, 
\citep{2018MNRAS.481.1707H} extended this analysis to the whole main-belt,
and, apart from identifying the case of the 329 Svea family, the third
asteroid family in terms of a population of asteroids in ${\nu}_6$ anti-aligned
states, found the first seven asteroids in aligned states of the same
resonance.

In this work, we are interested in further extending the analysis carried
out in previous works, by identifying more asteroids affected by the
${\nu}_6$ resonance.   According to \citet{Jones_2015}, the
catalogs produced by the Vera C. Rubin surveys will increase the known
number of tiny bodies in the Solar System by a factor of 10-100 times among
all populations, with $\simeq$ 1 million additional asteroid discoveries
projected in the first year of operations. New methods will be needed
to handle such a massive amount of data, and applications of machine
learning ($ML$) software development are currently under development
and will increase in the future.  The main goal of this paper is
to investigate the use of new methods for old problems, like image
identification and clustering methods.

For this purpose,  we first acquired the largest sample of asteroids'
proper elements for numbered and multi-opposition asteroids ever used
for this problem with methods
described in section~(\ref{sec: nu6_sel}).  Images of the time behavior of
asteroid resonant angles are then automatically identified using an
Artificial Neural Network ($ANN$) approach, developed by
\citet{2021MNRAS.504..692C} (section~(\ref{sec: ANN})).
The stability of the orbits in aligned and anti-aligned librating
states identified by the $ANN$ approach is then studied
when non-gravitational forces are considered in section~(\ref{sec: non-grav}).
The effect of local dynamics is then studied using dynamical maps
of synthetic proper elements in section~(\ref{sec: loc_dyn}).
Asteroid groups among the population of numbered asteroids in anti-aligned
libration states are then identified using  the Hierarchical Clustering Method
(HCM) \citep{2002aste.book..613B} in section~(\ref{sec: Ast_groups}).
Extension in domains of multi-opposition asteroids of the newly identified
groups and a study of the families' physical properties are discussed
in section~(\ref{sec: nu6_groups}).  In section~(\ref{sec: dat})
we date the newly discovered groups using methods based on time-reversal
numerical simulations.  Constraints on the initial ejection velocity field of
the 12988 Tiffanykapler family are discussed in
section~(\ref{sec: cons_quant}), while section~(\ref{sec: Tina_extra})
deals with explaining the albedo distribution and high eccentricity
population of the Tina asteroid family. Finally, the scientific importance
of the new database and the newly identified asteroid groups
are discussed in our conclusion, in section~(\ref{sec: concl}).

\section{Selecting the population of likely ${\nu}_6$ resonators}
\label{sec: nu6_sel}

To identify the population of ${\nu}_6$ resonators, first, we need
to select the asteroids most likely to be affected by this resonance.
For this purpose, we use the criterion described in \citep{2018MNRAS.481.1707H}:
we compute the value of $\sigma = {\varpi}-{\varpi}_6$ for all asteroids
of interest, and select those for which:

\begin{equation}
-3.5 \le \frac{d\sigma}{dt} \le 3.0 ~ arcsec/yr.
\label{eq: k6}
\end{equation}

\noindent  As discussed in \citep{2018MNRAS.481.1707H}, these are the bodies
that are more likely to be in aligned or anti-aligned librating states
of the resonance, according to the analytical model revised in
\citep{2018MNRAS.481.1707H}.  We applied this criterion to all the
numbered asteroids for which proper elements and frequencies are available
at the Asteroid Families Portal $AFP$
(``http://asteroids.matf.bg.ac.rs/fam/index.php'', \citet{Radovic_2017}),
i.e., those with an identification up to 523584, which yielded a sample
of 4875 asteroids likely to be affected\footnote{Synthetic proper
  elements as computed with the method of \citet{Knezevic_2003} are
  not very appropriate to study asteroids in secular resonances.  For
  these objects, resonant proper elements, as discussed in
  \citet{1993Icar..105...48M} and \citet{2011MNRAS.412.2040C}, should rather
  be used. However, since many of the asteroid families interacting with
  the ${\nu}_6$ secular resonances are crossed by it, and since
  resonant elements are not usually used for non-resonant asteroids,
  in this work we will use synthetic proper elements to
  identify resonant candidates and asteroid groups.  Past experiences
  showed that for family identification, despite
  differences in proper $e$ values, results obtained
  in domains of synthetic and resonant proper elements are fairly similar
  \citep{2011MNRAS.412.2040C, 2014ApJ...792...46C}.  The special case
  of the Tina asteroid family where the use of synthetic proper elements may
  overestimate the real value of proper $e$ will be treated in
  section~(\ref{sec: Tina_extra}).}.  

Osculating elements are available for an additional 83427 asteroids, up to
the identification of 607011 at the $AFP$.  For these objects, we first
computed proper elements using the \citet{Knezevic_2003} approach,
also described in \citet{2010MNRAS.408..580C}, and then applied
equation~(\ref{eq: k6}) to select the likely candidates.  This method
provided an additional sample of 1350 asteroids that can 
interact with the ${\nu}_6$ resonance.
Finally, $AFP$ also reports proper elements for a sample of 285861
multi-opposition asteroids.  By applying our selection criteria, we
obtained an additional population of 3779 asteroids likely to interact
with the ${\nu}_6$ resonance.

\section{$ANN$ identification of resonant orbits}
\label{sec: ANN}

The whole sample of asteroids potentially affected by the ${\nu}_6$ resonance
is 10004.  To identify the population of aligned and antialigned
asteroids, we first used the approach described in \citet{2021MNRAS.504..692C},
which is a five-step process:  first, we integrate the asteroid orbits 
under the gravitational influences of all planets, then we compute
the time-series of the ${\nu}_6$ resonant argument, images of these
time-series are obtained for each asteroid.   The $ANN$ model
is trained on a training set of labeled image data and, finally, predictions
of the labels for a set of test images, usually 50 at a time, are obtained
and confirmed by visual inspection by all authors independently.

For this task, we used the $ANN$ model described in
\citet{2021MNRAS.504..692C}, which is a four-layer neural network
with a flatten, inner, hidden, and output layers.  Interested
readers can find more information in the cited paper.  An example of
the outcome of this approach is shown in figure~(7)
of \citet{2021MNRAS.504..692C}.
The application of these methods lead to the identification of 30
asteroids in aligned resonant configurations, 9 of which are among
the multi-opposition population, and 1713 asteroids in anti-aligned
configurations, 667 of which among the multi-opposition asteroids.
The long-term stability of these orbits when non-gravitational forces
are accounted for will be discussed in the next section.

\section{Effects of non-gravitational forces}
\label{sec: non-grav}

So far, we analyzed the asteroids' orbits in a purely conservative scenario.
Here, we investigate the long-term stability of the objects found in
aligned and antialigned states when non-gravitational forces like the
Yarkovsky effect are considered.  For this purpose, we integrate the
population of librating asteroids with $SWIFT-RMVSY$, a symplectic
integrator modified to account for the diurnal and seasonal version
of the Yarkovsky effect \citep{1999MsT..........2B}.  We created
two sets of clones of these objects, one with a spin obliquity of
$+90^{\circ}$ and one with $-90^{\circ}$.  The prograde particles will
maximize the Yarkovsky drift and move towards larger values of semi-major
axis, while the retrograde particles will move toward lower values of $a$
at the maximum possible rate.  We neglected in this simulation the YORP
effect, which could alter the spin obliquity and asteroid period,
and collisions, which could also affect these quantities.
We use sizes obtained from the asteroid absolute magnitudes assuming
an albedo $p_V$ of 0.12, which divides C-complex from S-complex objects
and can be assumed to be a reasonable compromise between the two.
For the other key parameters of the Yarkovsky force, we used a density
and a bulk density of 1500 $kg m^{-3}$, a thermal conductivity $K$
of 0.01 W/(m K), a specific heat $C_P = 680 J kg^{-1} K^{-1}$
a Bond albedo of 0.05, and an infrared emissivity $\epsilon$ of 0.95.
The test particles were integrated over 5 My over the gravitational
influence of all planets and the time behavior of the ${\nu}_6$
resonant angle was verified for both clones.

\begin{figure*}
  \centering
    \begin{minipage}[c]{0.45\textwidth}
    \centering \includegraphics[width=2.3in]{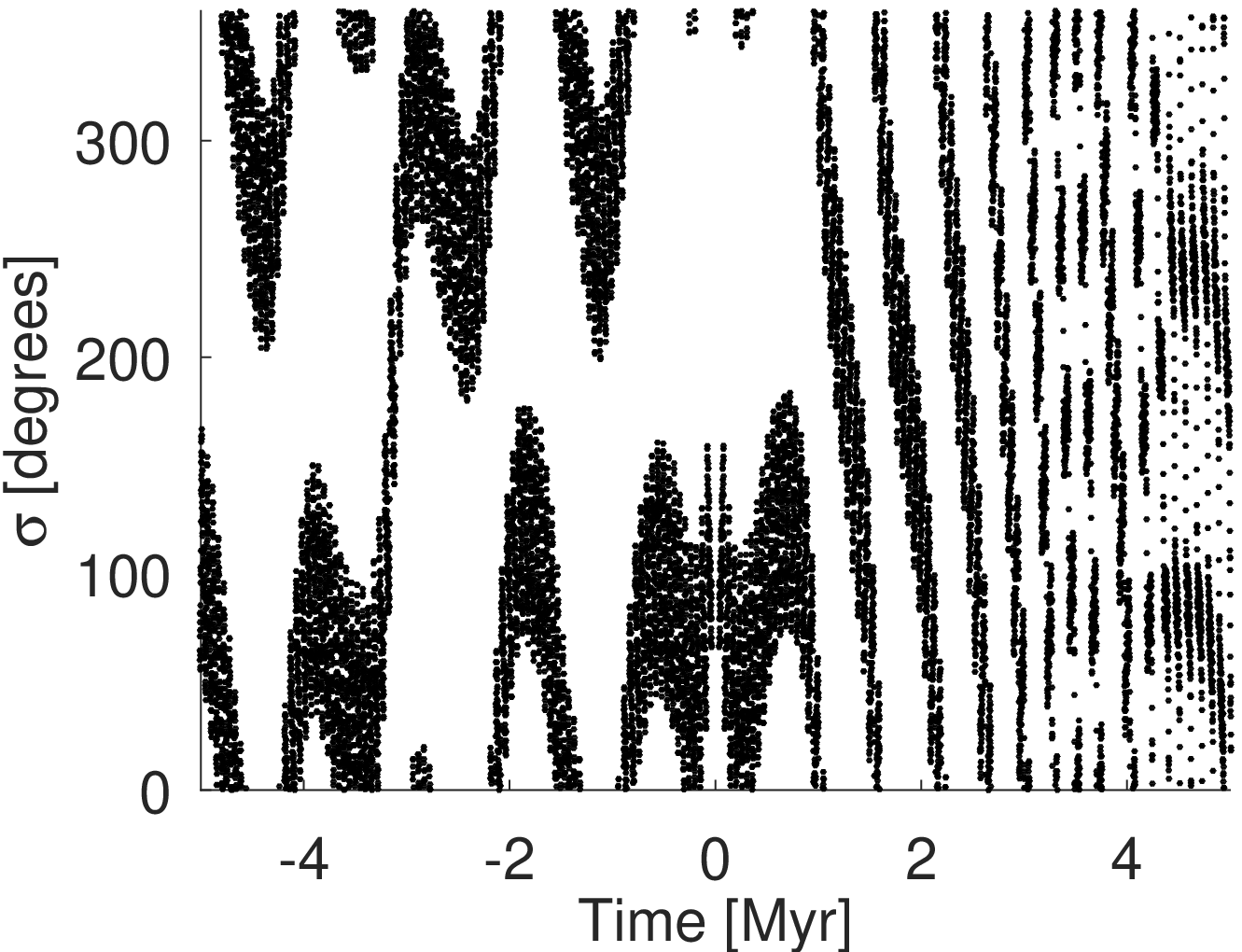}
  \end{minipage}%
  \begin{minipage}[c]{0.45\textwidth}
    \centering \includegraphics[width=2.3in]{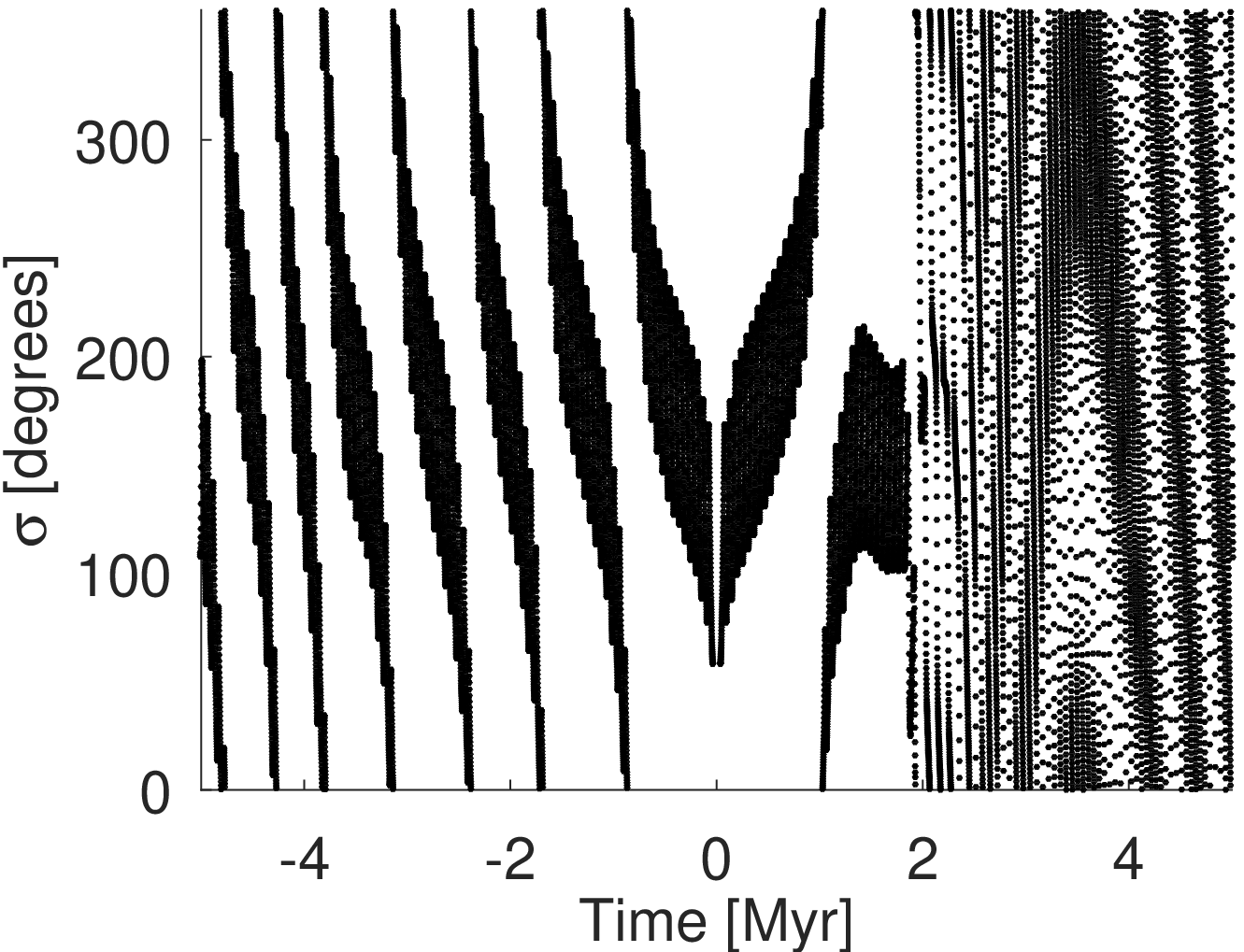}
  \end{minipage}

  \caption{Time behavior of the ${\nu}_6$ resonant angle for two clones of
    asteroids that escaped from aligned configuration
    (asteroid 538256, left panel) and anti-aligned configuration (asteroid
    60683, right panel).  The time interval from -5 My to 0 refers to the
    evolution of the retrograde clone, while the interval from 0 to 5 My
    displays the behavior of the prograde one.}
\label{fig: nu6_rmvsy}
\end{figure*}

If both clones remained in a librating configuration over the length of
the simulation, the librating behavior was deemed as confirmed.  However,
if one of the clones escaped from aligned or antialigned libration during
the simulation, we dropped the corresponding asteroids
from our list of resonant objects.  Figure~(\ref{fig: nu6_rmvsy}) displays
the time behavior of the ${\nu}_6$ resonant angle for particles
that escaped aligned libration (left panel) and anti-aligned libration
(right panel) because of the Yarkovsky force.

\begin{figure}
  \centering
  \centering \includegraphics[width=3.5in]{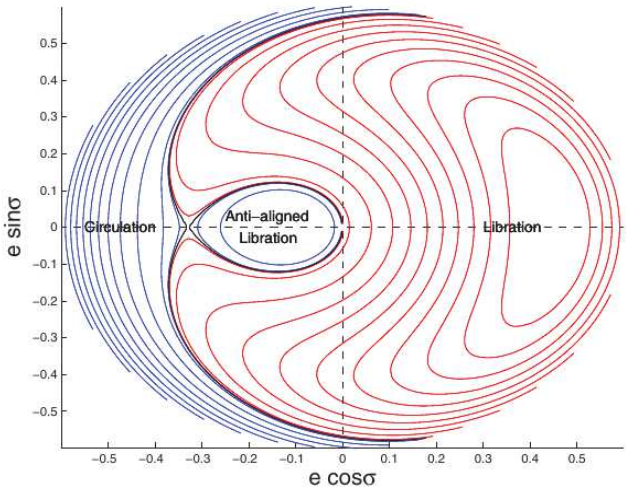}
  \caption{A diagram of equi-Hamiltonian levels for the orbital
    region of the Tina family.  Circulating orbits are shown in blue,
    while librating ones are in red.  The resonance separatrix is
    shown as a black line. Adapted from figure~(1) of
    \citet{2011MNRAS.412.2040C}.}
\label{Fig: Tina_Hamiltonian}
\end{figure}

Out of 30 asteroids in possible aligned configurations, only 15
passed this test, and only one of them was a multi-opposition asteroid.
The survival rate was much higher for antialigned orbits: out of 1713 asteroids,
only 44 escaped the antialigned configuration because of non-gravitational
forces, i.e., 2.57\%.  This difference in behavior between
aligned and antialigned orbits can be better understood if we consider
the outcome of analytical models of the ${\nu}_6$ resonance
\citep{1991CeMDA..51..169M}.  Figure~(\ref{Fig: Tina_Hamiltonian})
displays the energy levels of the Hamiltonian model of
\citet{1991CeMDA..51..169M}, computed for the orbital region of
the Tina family.  Interested readers could find more details on
how this figure was obtained in \citet{2011MNRAS.412.2040C}.
The black line separates circulating orbits from librating ones and cannot
be crossed in the analytical approximation.
Anti-aligned libration orbits are truly circulating ones, captured
in a loop of the separatrix, and this explains their higher rate of survival.
An antialigned orbit has to cross the resonance separatrix to become
a circulating orbit and be lost because of planetary close encounters.
Stable aligned orbits are just the red librating orbits of the figure
with small oscillating amplitudes.  An increase
in the libration amplitude caused by planetary perturbations is enough
to destabilize the aligned configuration.  This explains why aligned
orbits are rarer and more unstable when non-gravitational forces are
considered.
The next section will focus on better understanding
the effects of the local dynamics on the confirmed population of stable
${\nu}_6$ librators.

\section{Local Dynamics}
\label{sec: loc_dyn}

The dynamical location of the ${\nu}_6$ resonance has been discussed
in detail in several prior papers.  Interested readers could find
more information in \citet{2018MNRAS.481.1707H}.  Here, we briefly revise
some basic concepts, using dynamical maps of synthetic proper elements
to highlight the local dynamical environment.

Synthetic proper elements are quasi-integral of motions obtained
through Fourier analysis of time-series of osculating elements
\citep{Knezevic_2003}.  Here we apply the procedure described
in \citet{2010MNRAS.408..580C} to obtain dynamical maps of
synthetic proper elements.  Essentially, a grid of initial conditions
in a 2-dimensional plane is created, and the test particles are integrated
over the gravitational influence of all planets with a symplectic
integrator, like, for instance, the $SWIFT\_MVSF$ mixed-variable
symplectic integrator of \citet{1994Icar..108...18L}, modified
by \citet{1999MsT..........2B} to allow for online filtering.
The four elements not in the bi-dimensional grid are usually taken
equal to those of an asteroid of interest, in our case 1222 Tina.
The initial conditions and displacements in orbital elements for our maps
are shown in table~(\ref{Table: Dyn_maps}).  All maps had 35 initial
conditions on the x-axis and 60 on the y-axis, for a total of 2100
test particles.
   
   \begin{table*}
      \begin{center}
        \caption{For each of the six maps shown in figure~(\ref{fig: nu6maps}),
          we report the initial value of the element used in the x-axis, in
          the y-axis, and the displacement used for the grids in x and y.}
        \label{Table: Dyn_maps}
         \begin{tabular}{|c|c|c|c|c|}
\hline
Map     & Initial & Initial & Change in & Change in \\
Id.     & el. (1) & el. (2) & el. (1)   & el. (2)  \\  
\hline
1 &  2.4850 [au] & 16 [degrees] & 0.0100 [au] & 0.125 [degrees] \\
2 &  2.4850 [au] & 0            & 0.0100 [au] & 0.005          \\
3 &  2.0800 [au] & 0 [degrees]  & 0.0115 [au] & 0.3 [degrees]  \\
4 &  2.0800 [au] & 0            & 0.0115 [au] & 0.005          \\
5 &  2.8258 [au] & 19 [degrees] & 0.0107 [au] & 0.21 [degrees] \\ 
6 &  2.8258 [au] & 0            & 0.0107 [au] & 0.0061         \\
\hline
\end{tabular}
\end{center}
\end{table*}

\begin{figure*}
  \centering
  \begin{minipage}[c]{0.45\textwidth}
    \centering \includegraphics[width=2.3in]{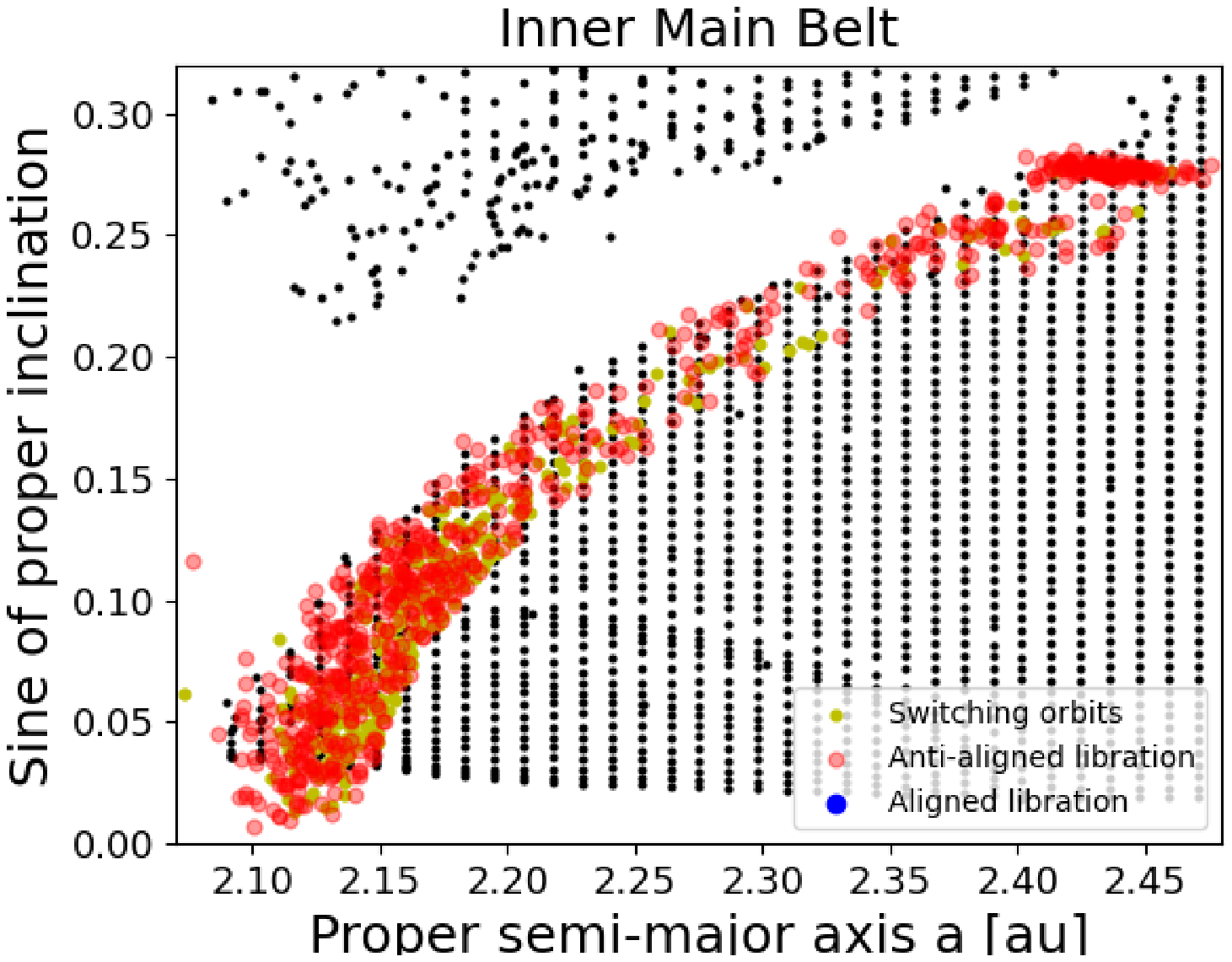}
  \end{minipage}%
  \begin{minipage}[c]{0.45\textwidth}
    \centering \includegraphics[width=2.3in]{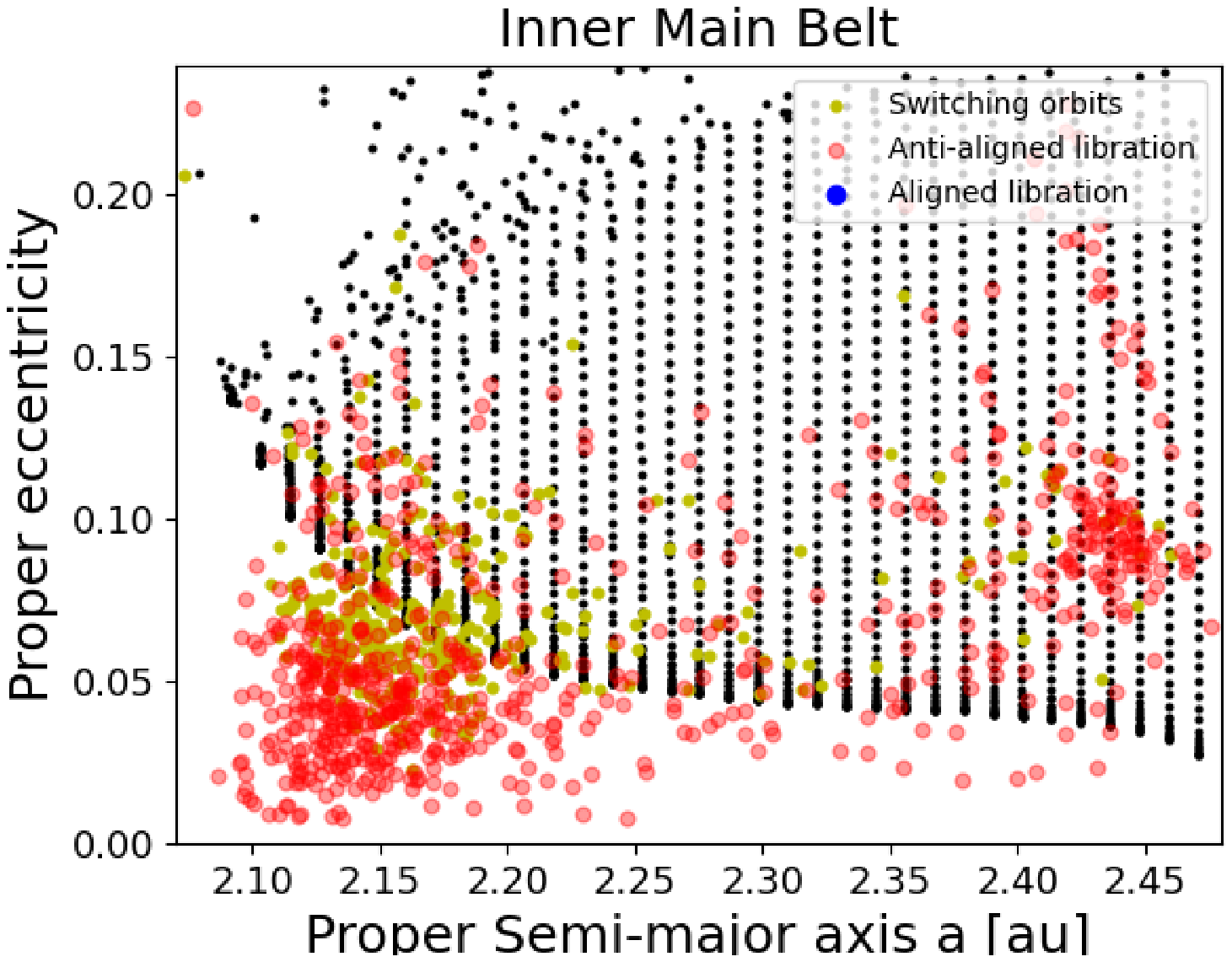}
  \end{minipage}
  \begin{minipage}[c]{0.45\textwidth}
    \centering \includegraphics[width=2.3in]{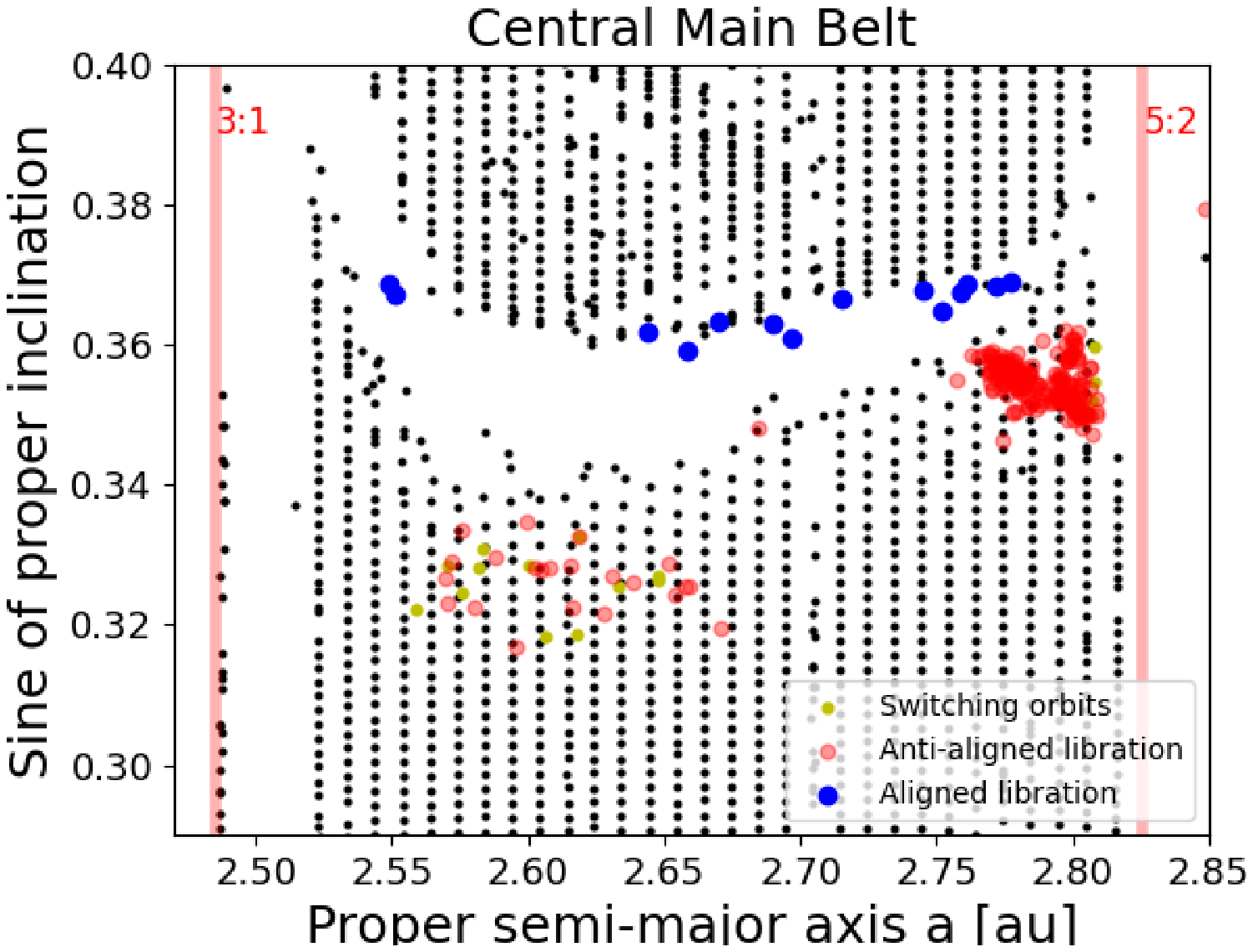}
  \end{minipage}%
  \begin{minipage}[c]{0.45\textwidth}
    \centering \includegraphics[width=2.3in]{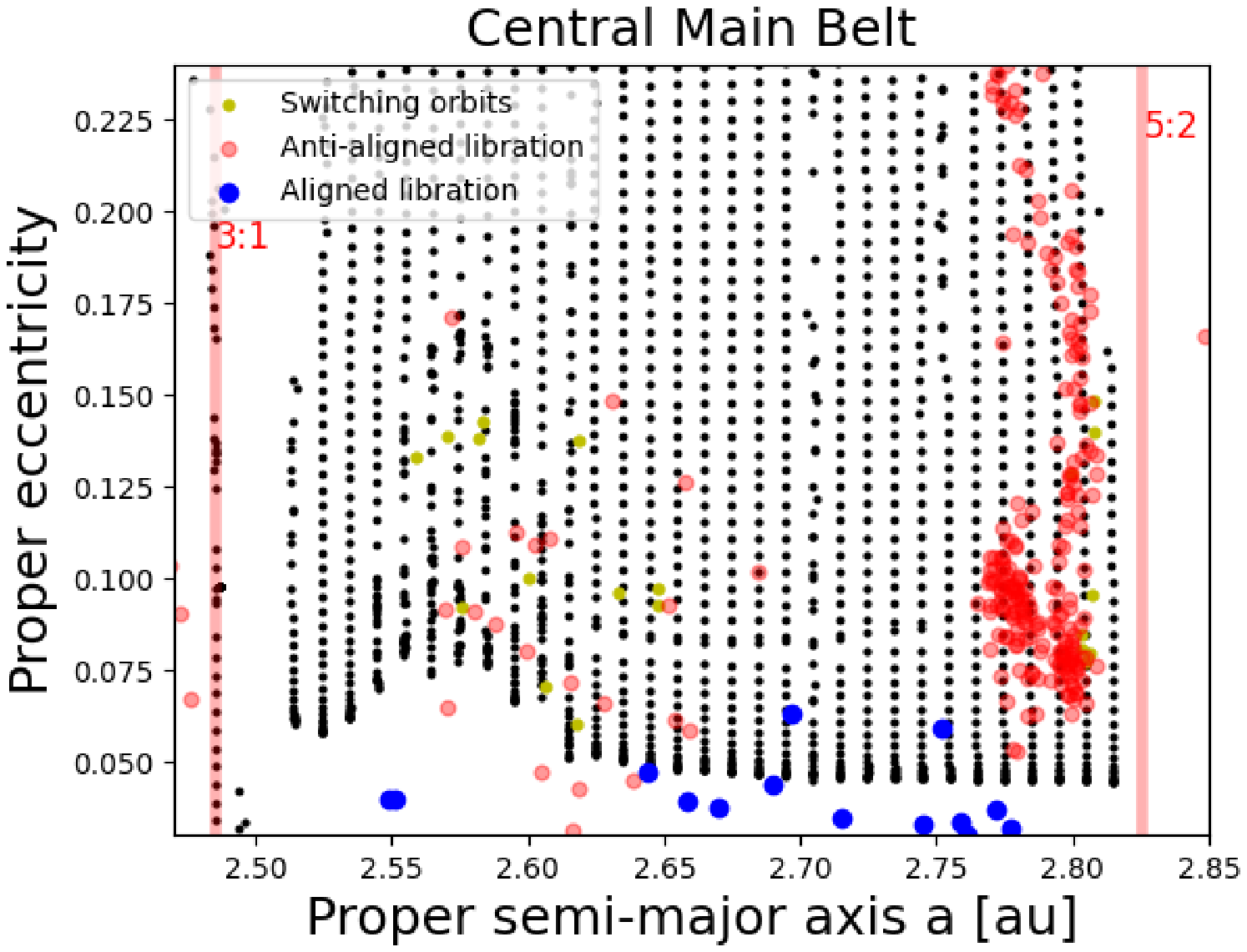}
  \end{minipage}
  \begin{minipage}[c]{0.45\textwidth}
    \centering \includegraphics[width=2.3in]{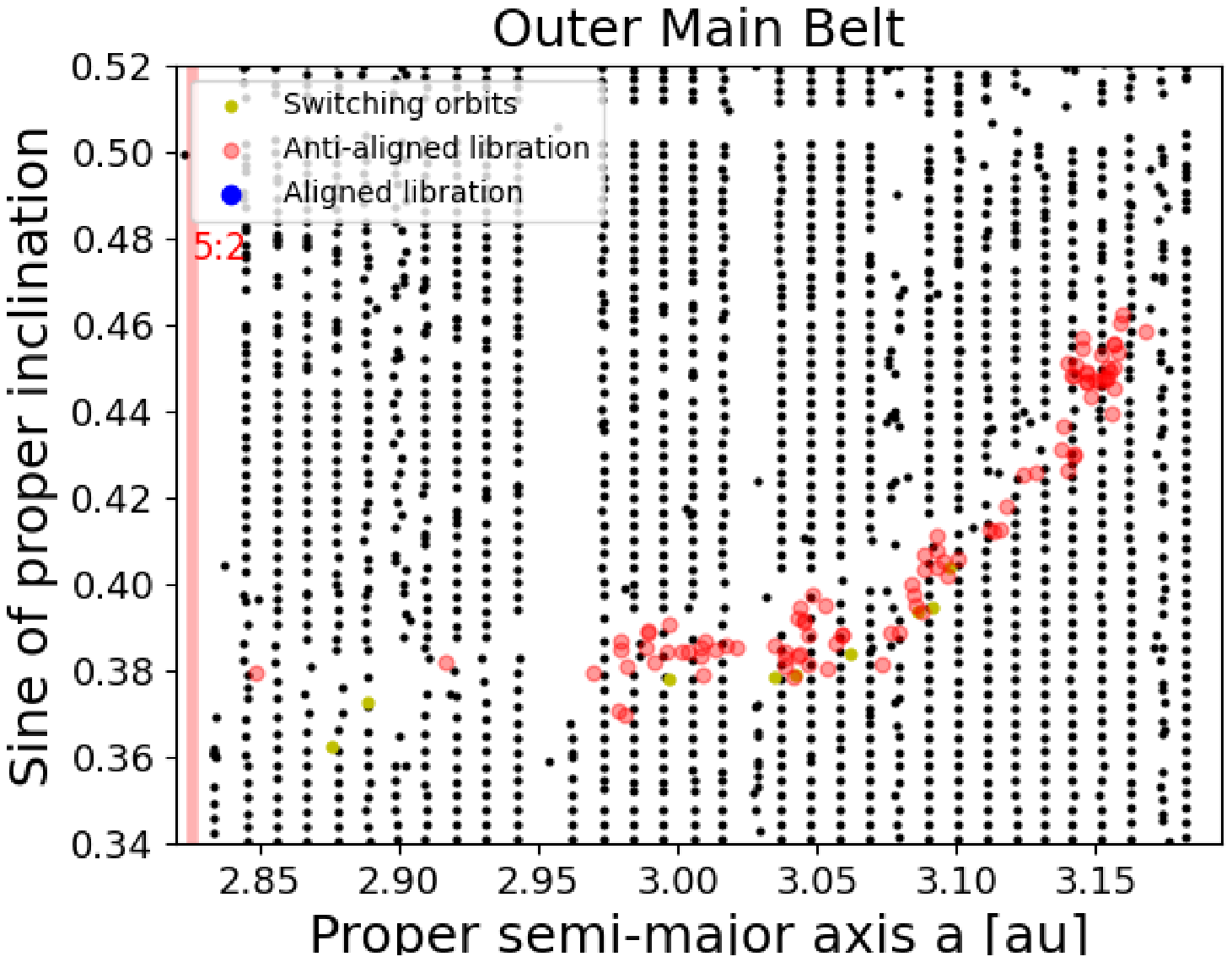}
  \end{minipage}%
  \begin{minipage}[c]{0.45\textwidth}
    \centering \includegraphics[width=2.3in]{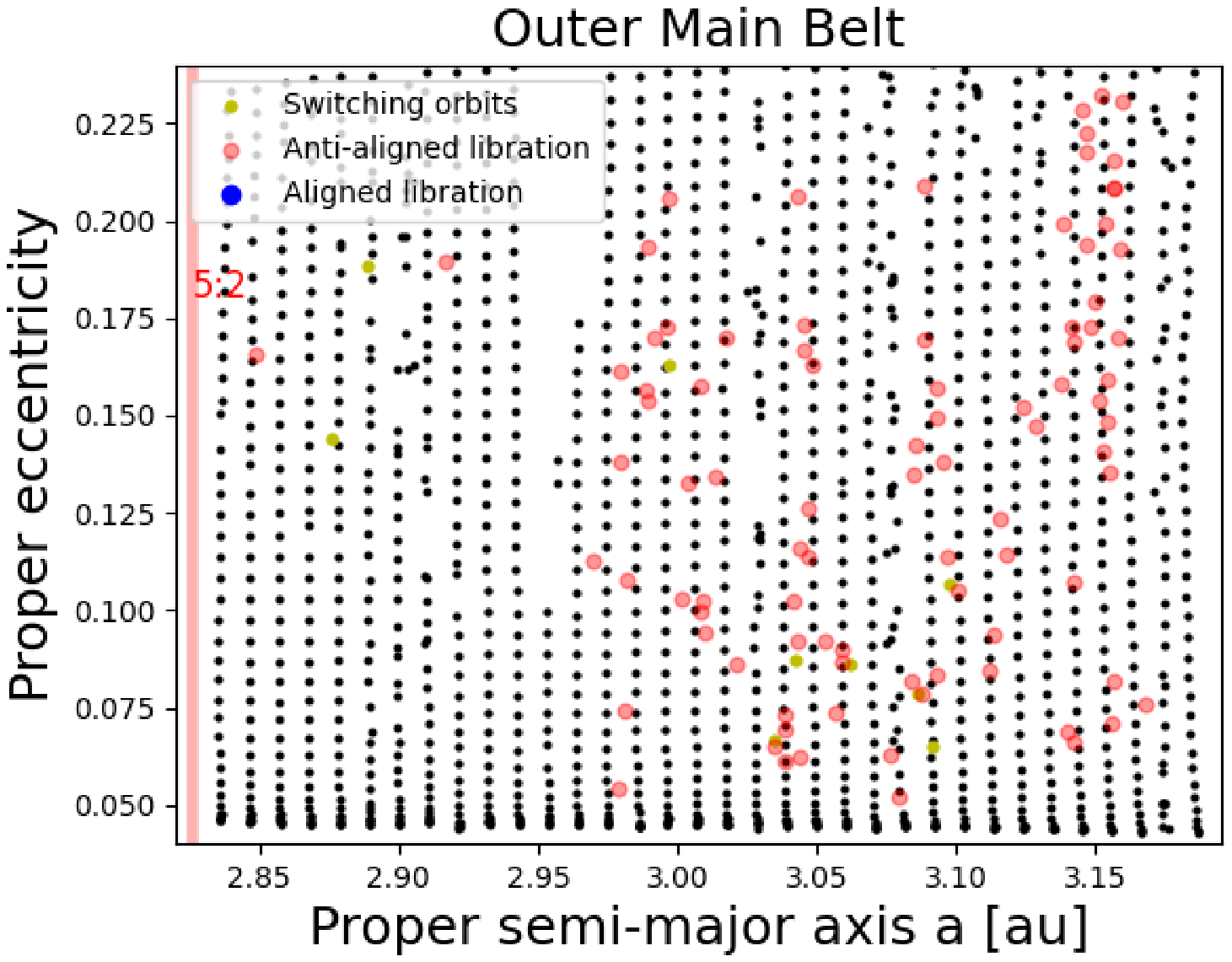}
  \end{minipage}  
  \caption{Dynamical maps of synthetic proper
    elements in the $(a,\sin{(i)})$ plane (left panels), and in the
    $(a,e)$ plane (right panels).  The top panels display asteroids
    in the inner main-belt, the central panels those in the central
    main-belt, and the bottom panels asteroids in the outer main-belt.
    Black dots show the orbital location of proper elements
    for the test particles used in the six simulations.  Red full
    circles identify real asteroids in anti-aligned libration states,
    blue full circles real asteroids in aligned libration states, and
    yellow full circles asteroids that alternate phases of libration
    and circulation.}
\label{fig: nu6maps}
\end{figure*}

Here we created six maps in the $(a, \sin(i))$, and $(a,e)$ planes,
for the inner, central, and outer main-belt. The initial conditions and
number of particles for our maps are shown in table~(\ref{Table: Dyn_maps}).
Our results are shown in figure~(\ref{fig: nu6maps}).  Synthetic proper
elements for the
test particles are shown as black full circles.  The populations
of numbered asteroids that are on aligned, anti-aligned librating or
switching particles (15, 1046, and 362, respectively) are shown as blue, red
and yellow full circles.
Mean-motion resonances appear as vertical bands with low number-density of
asteroids, while secular resonances are inclined bands in the
$(a, \sin(i))$ and $(a,e)$ domains. The lack of test particles at low
eccentricities is an artifact of the process used to generate the
dynamical maps.

In the inner main-belt, we have two main agglomerations, one at 2.45 au
associated with the Svea asteroid family, and another at lower values
of proper $a$ that follows the lower contours of the unstable region caused
by the ${\nu}_6$ secular resonance.  In the central main-belt, we have
the group associated with the 1222 Tina family, scattered asteroids
at lower $a$, and a population of objects in aligned libration
states that follow the upper boundary of the ${\nu}_6$ unstable region.
In the outer main-belt, we have the group of anti-aligned asteroids
associated with the Euphrosyne family.

\begin{figure}
  \centering
  \centering \includegraphics[width=3.5in]{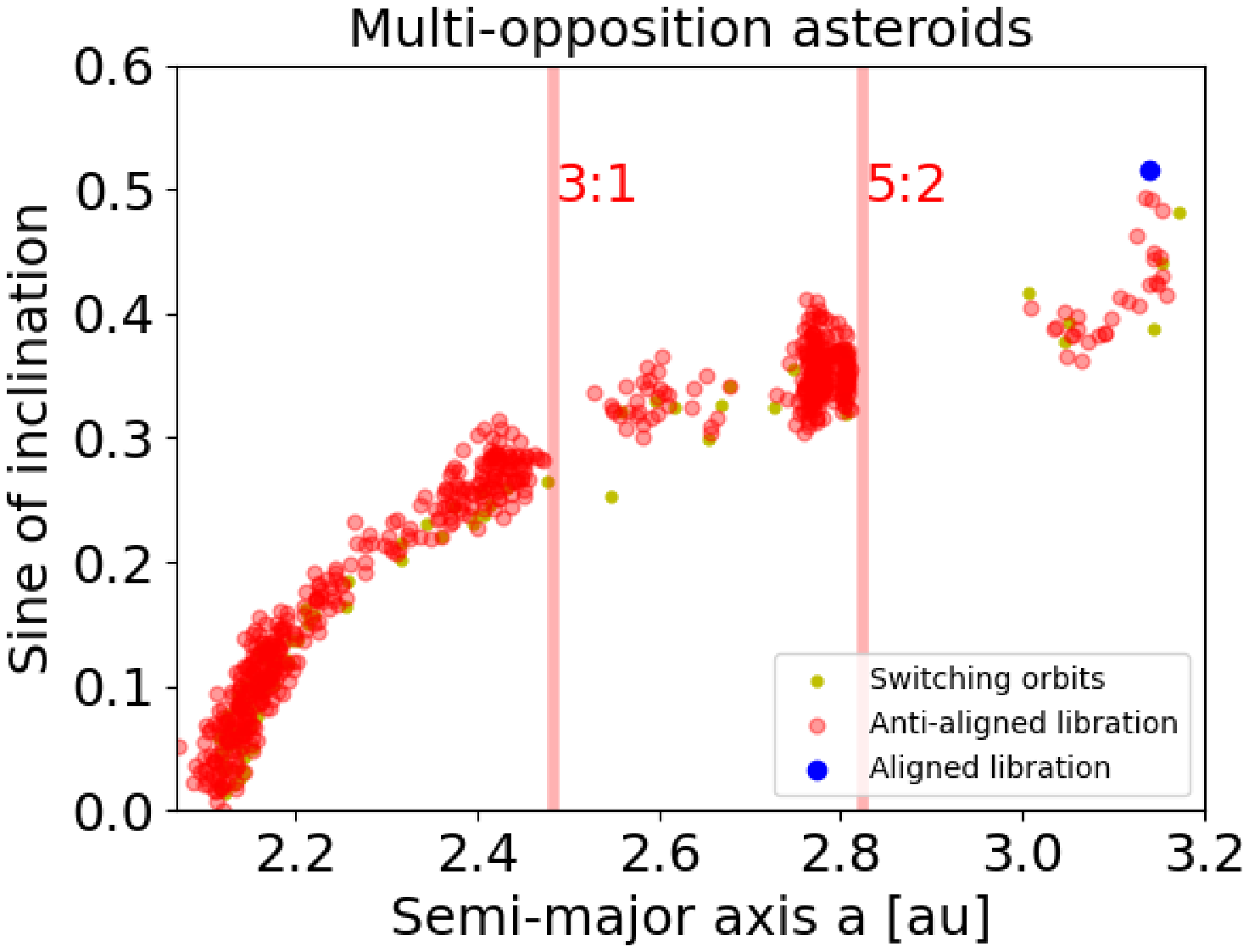}
  \caption{An $(a,\sin(i))$ projection of multi-opposition
    asteroids (right panel) in aligned, anti-aligned and switching states
    of the ${\nu}_6$ secular resonance.  The symbols have the same meaning
    as in figure~(\ref{fig: nu6maps}).}
\label{Fig: multi_opp}
\end{figure}

If we also consider the multi-opposition asteroids, then we have 9
more aligned, 667 antialigned, and 90 switching objects.  Our results
are displayed on figure~(\ref{Fig: multi_opp}). Trends previously observed
for the numbered objects are mostly confirmed by the multi-opposition
population as well, except for aligned asteroids. We only
found one multi-opposition asteroid in the outer main-belt,
near the 2:1 resonance, whose orbit was stable when the Yarkovsky force
was considered: 2015 DP197. In the next section, we aim to identify
groups among the antialigned asteroidal population that has been detected so
far.

\section{Asteroid groups}
\label{sec: Ast_groups}

The Hierarchical Clustering Method (HCM) has been used since the early 90s
to identify asteroid families \citep{2002aste.book..613B}.  The most
common variant of the method will look for neighbors of an asteroid
in a domain of the $(a,e,\sin{(i)})$ proper elements, using a distance metric
appropriate for the problem at hand.   If the distance between the first
asteroid and the second is less than a critical value, $d_0$, the second
asteroid is assigned to the family membership list and the procedure
is then repeated for the second body, until no new members are encountered.
$d_0$ is a free parameter of the method, that depends on the number
density of asteroids in a given orbital region.  Here we follow the approach of
\citet{2001Icar..153..391B} that defines $d_0$ as the mean of
the minimum distances between all asteroids in the region.

Two useful methods for identifying asteroid families are obtaining
family membership versus cutoff and stalactite plots.  For the first
method, one plots the number of family members as a function of
increasing distances, up to values of distances high enough that 
all asteroids in the orbital region are associated with a given
family.  In a stalactite diagram, as constructed by \citet{2008MNRAS.390..715B},
first all the asteroids in a family that includes all the objects
in the region are displayed.  The cutoff distance value is then
reduced and asteroid groups are identified among the population
of objects no longer associated with the initial family.
The procedure is then repeated for lower and lower values of
the distance cutoff.  Results are presented in a plane where the
families' memberships are shown on the x-axis and the distance
cutoff is shown on the y-axis.   More details on all
these methods and procedures are described in \citet{2010MNRAS.408..580C}.
  
\begin{table*}
  \begin{center}
    \caption{Number of asteroids on antialigned orbits
      and cutoff distance $d_0$ for the inner, central and outer main
      belt regions.}
    \label{Table: MB_regions}
    \begin{tabular}{|c|c|c|}
\hline
Main-Belt & Number of & Cutoff distance\\
region    & asteroids &  $d_0 [m/s]$ \\
\hline
Inner   & 682 & 160.5 \\
Central & 252 &  69.7 \\
Outer   & 112 & 185.5 \\
\hline
\end{tabular}
\end{center}
\end{table*}
   
Our dataset of proper elements is comprised of the numbered
asteroids in antialigned states of the ${\nu}_6$ resonance.
Except for a single object with a semi-major axis beyond
the 2:1 mean-motion resonance with Jupiter, in the Cybele
region, all the other asteroids can be found in the
inner main-belt, for values of $a$ lower than those of the
3:1 mean-motion resonance with Jupiter ($a_{3:1} = 2.485$ au),
in the central main-belt ($ a_{3:1} < a < a_{5:2}$, with $a_{5:2} = 2.826$ au),
and in the outer main-belt ($a_{5:2} < a < a_{2:1}$, where $a_{2:1} = 3.279$ au).
Table~(\ref{Table: MB_regions}) displays the number of asteroids and the
values of $d_0$ for the three regions of interest.  We define a family as a
group with at least 5 members.  In the next section, we will discuss our
results for the inner main-belt.

\subsection{Inner main-belt}
\label{sec: inn_belt}

The inner main-belt contains the largest population of asteroids in
${\nu}_6$ antialigned resonant states.  Two main groups of
asteroids are found in this region, one for $a < 2.30 $ au, where most
of the asteroidal population is found, and one for higher values,
mostly associated with the Svea family, as discussed in
\citet{2018MNRAS.481.1707H}.

\begin{figure*}
  \centering
  \begin{minipage}[c]{0.45\textwidth}
    \centering \includegraphics[width=2.3in]{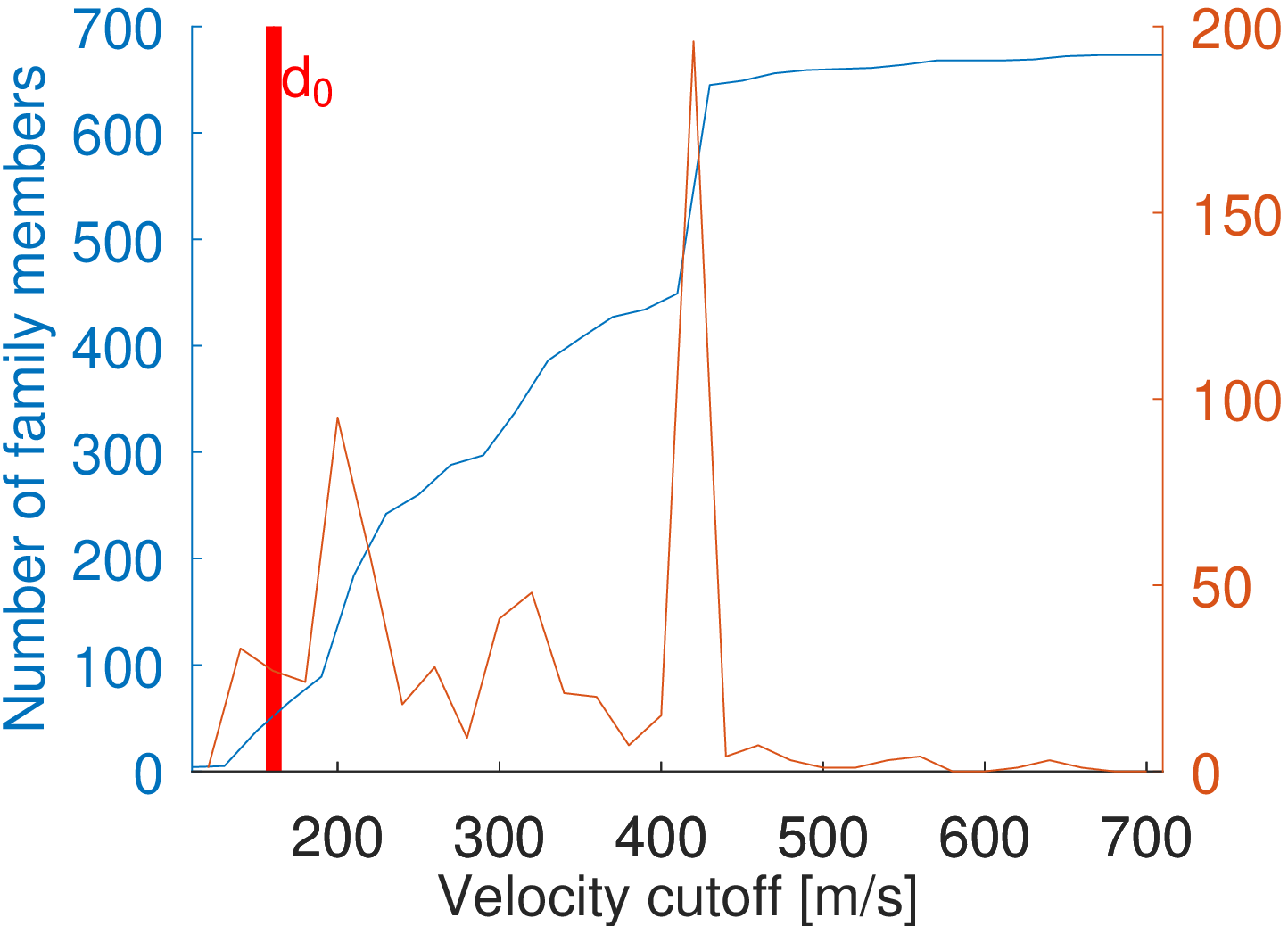}
  \end{minipage}%
  \begin{minipage}[c]{0.45\textwidth}
    \centering \includegraphics[width=2.3in]{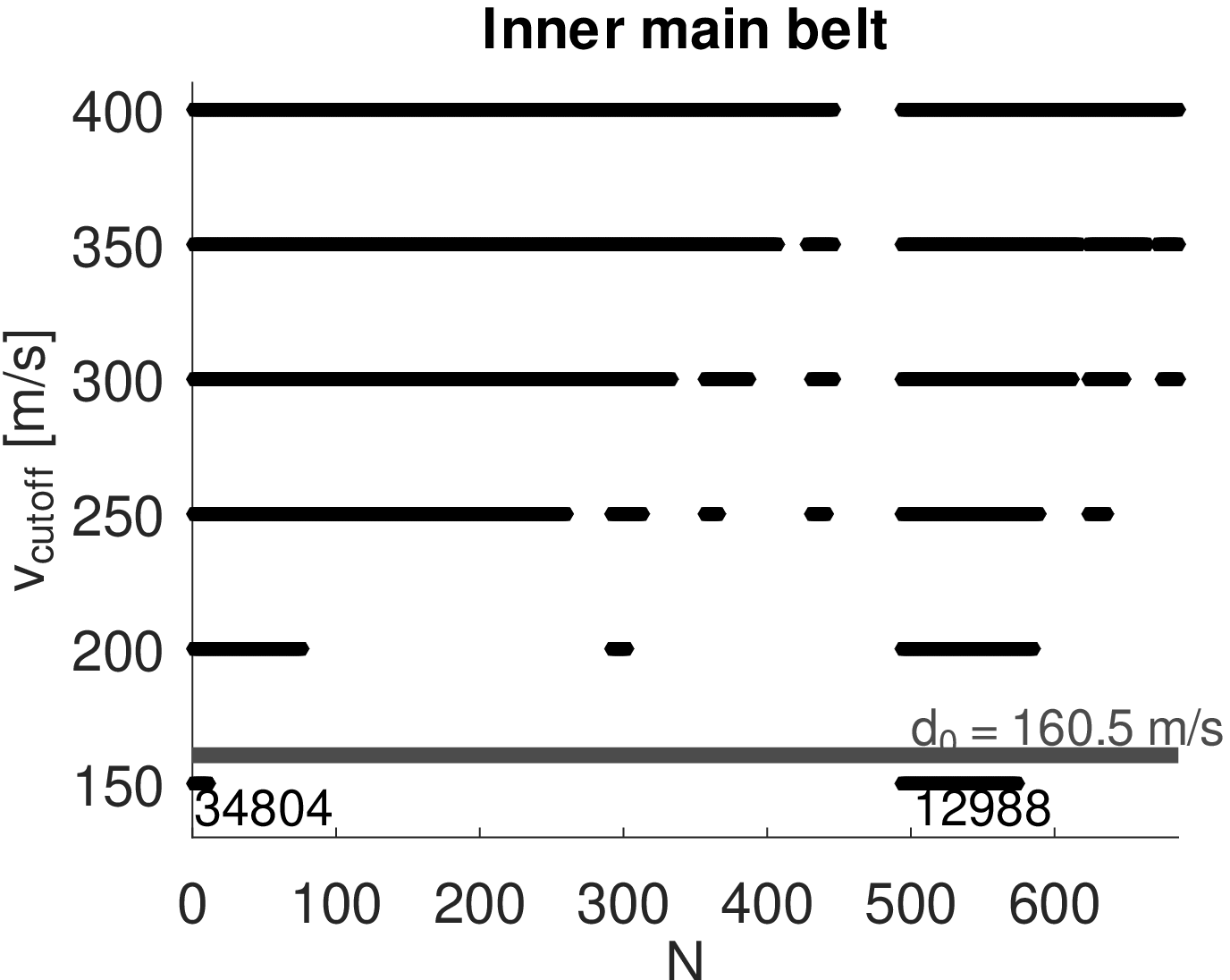}
  \end{minipage}
  \caption{On the left panel we show the number of members of the group
    associated with the lowest  numbered asteroid in the
    inner main-belt, 930 Westphalia, as
    a function of the distance cutoff (blue line).  The orange line shows the
    differential increase in the group members for each increase in the cutoff
    value.  The thick red line displays the critical cutoff value for the
    inner main-belt.  On the right panel, we display a stalactite diagram for
    the same region.  Black full dots identify members of asteroid groups
    at various values of the distance cutoff.  The critical value for this
    region is shown as a thick horizontal line.  The identified clusters
    are identified by the lowest id number among their members.}
\label{fig: groups_inner}
\end{figure*}

The left panel of figure~(\ref{fig: groups_inner}) shows a family count plot
for groups starting with the Svea family members.  As we increase
the distance cutoff, several smaller groups are englobed by the main
group, as shown by the peaks in the differential increase in the
group members (orange line).  At $d_0 = 420$ m/s the Svea family
merges with the rest of the population, and from then on essentially all
asteroids are linked in a unique group.
The stalactite diagram for this region identifies two main groups:
a group around 34804 2001 SP67, associated with the Svea family,
and a completely new group in the inner main-belt around 12988 Tiffanykapler.
As discussed by \citet{2018MNRAS.481.1707H}, the Svea family is cut
by the ${\nu}_6$ secular resonance in two parts and the part near asteroid
329 Svea itself is not in a resonant configuration.  This explains
why the Svea group's lowest-numbered member is 34804 and not 329 Svea.
The new Tiffanykapler group is located at $a \simeq 2.182$ au and it is not
part of any currently known family.  The dynamical and physical properties
of this new group will be discussed in section~(\ref{sec: nu6_groups}).

\subsection{Central main-belt}
\label{sec: cen_belt}

\begin{figure*}
  \centering
  \begin{minipage}[c]{0.45\textwidth}
    \centering \includegraphics[width=2.3in]{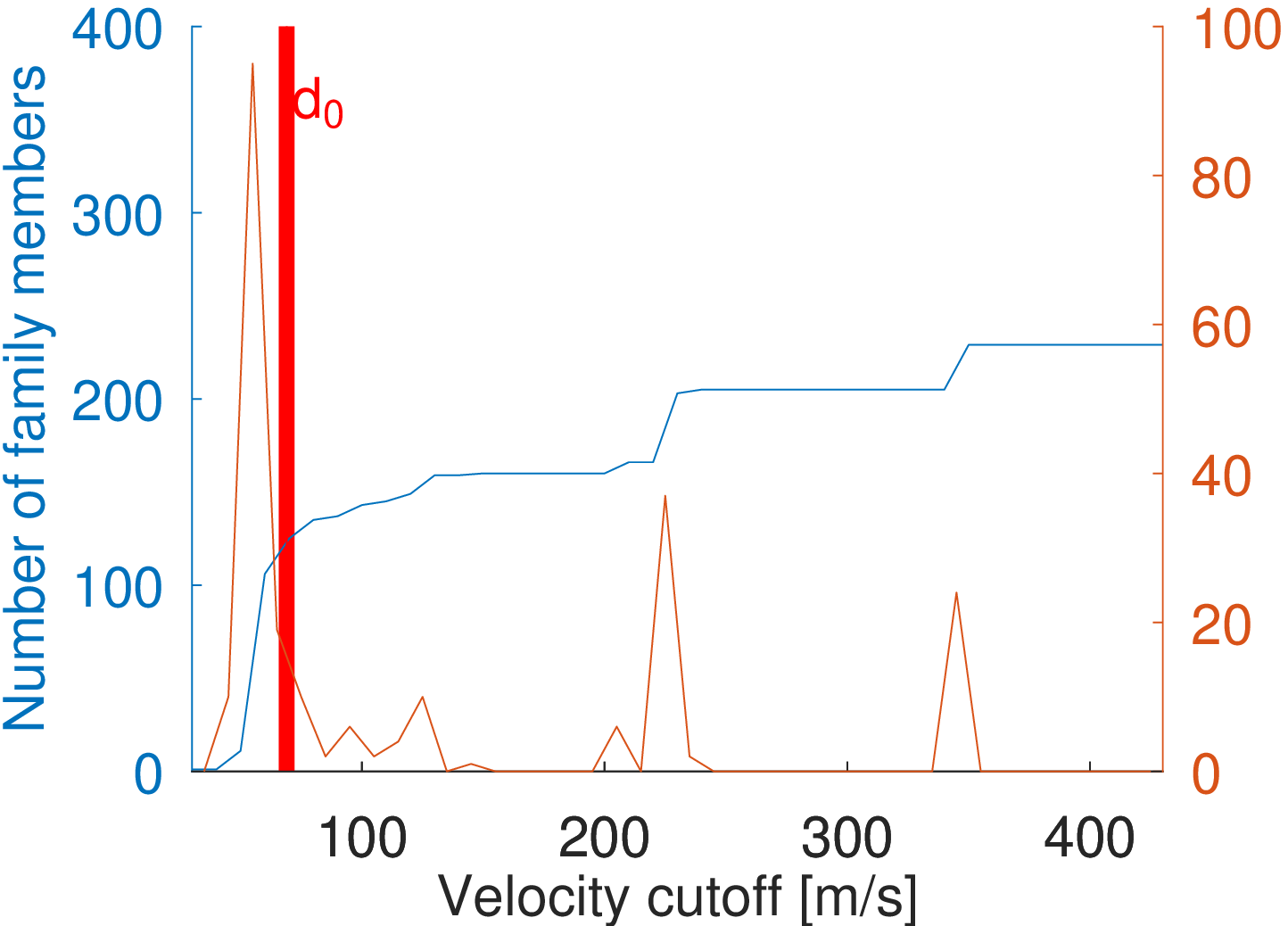}
  \end{minipage}%
  \begin{minipage}[c]{0.45\textwidth}
    \centering \includegraphics[width=2.3in]{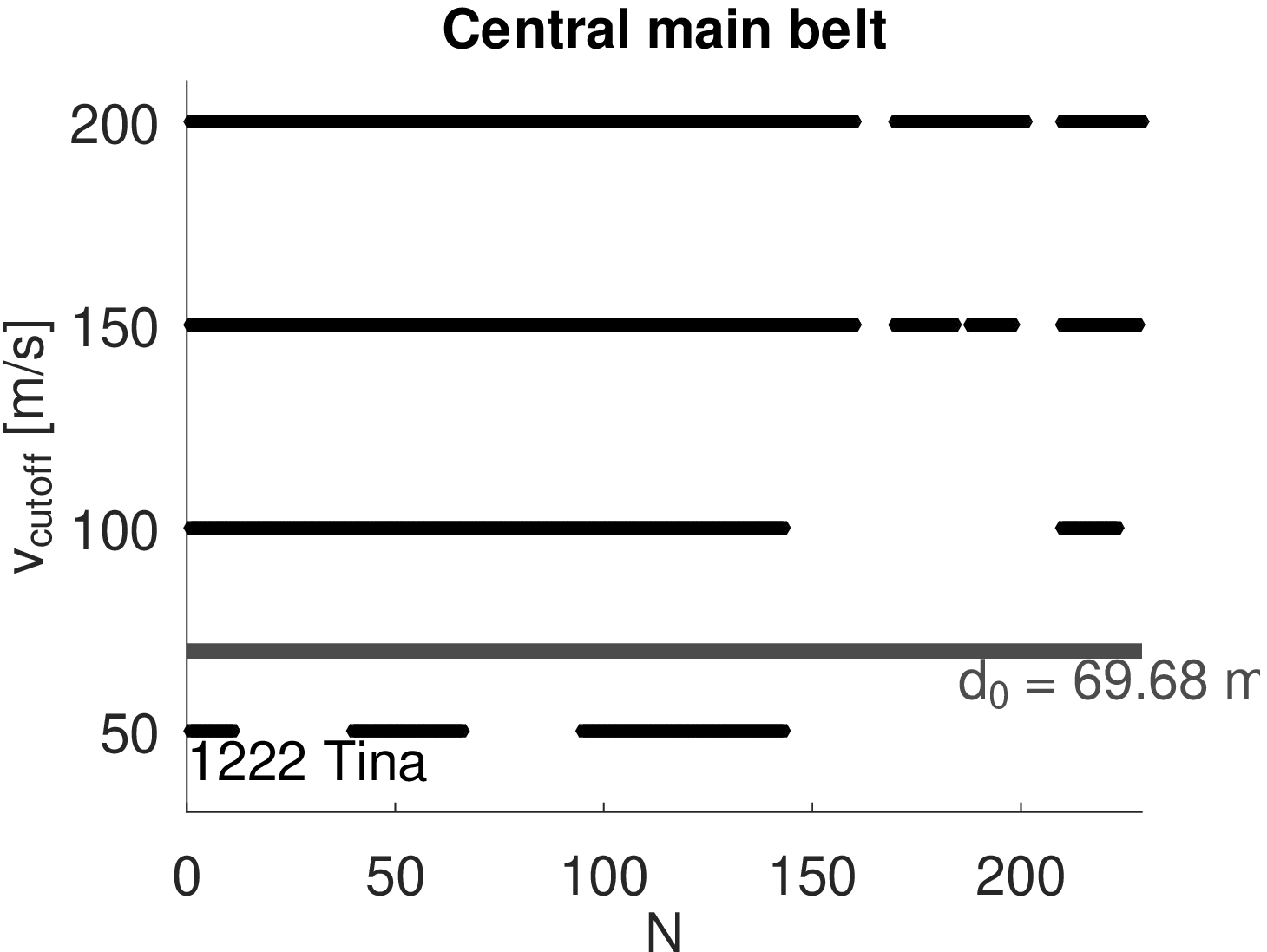}
  \end{minipage}
  \caption{The same as in figure (\ref{fig: groups_inner}), but for the central
    main-belt.  We used 1222 Tina as a possible parent body to generate
    both panels.}
  \label{fig: groups_central}
\end{figure*}

The main asteroid group in the central main-belt is that of 1222 Tina,
and no other group has been identified so far in this region
\citep{2011MNRAS.412.2040C}.  This picture is confirmed by the
analysis performed in this work.  The Tina family essentially
coalesces for cutoff values slightly lower than the local
distance cutoff $d_0 = 69.7$ m/s, and only minor groups are
merging with the main family at larger cutoffs (see figure~(\ref{fig: groups_central}, left panel)).  The stalactite diagram confirms this analysis and
at the distance cutoff computed for the central main-belt
only the Tina family is observed as a group.
Concerning previous determinations of the Tina family, however,
we identify in this work a population of objects at higher eccentricities
that are not members of the Tina dynamical family (see
figure~(\ref{fig: nu6maps}, center right panel).  No single
dynamical group has been identified among this new population,
and its nature will be investigated in more detail in
section~(\ref{sec: Tina_extra}).

\subsection{Outer main-belt}
\label{sec: out_belt}

\begin{figure*}
  \centering
  \begin{minipage}[c]{0.45\textwidth}
    \centering \includegraphics[width=2.3in]{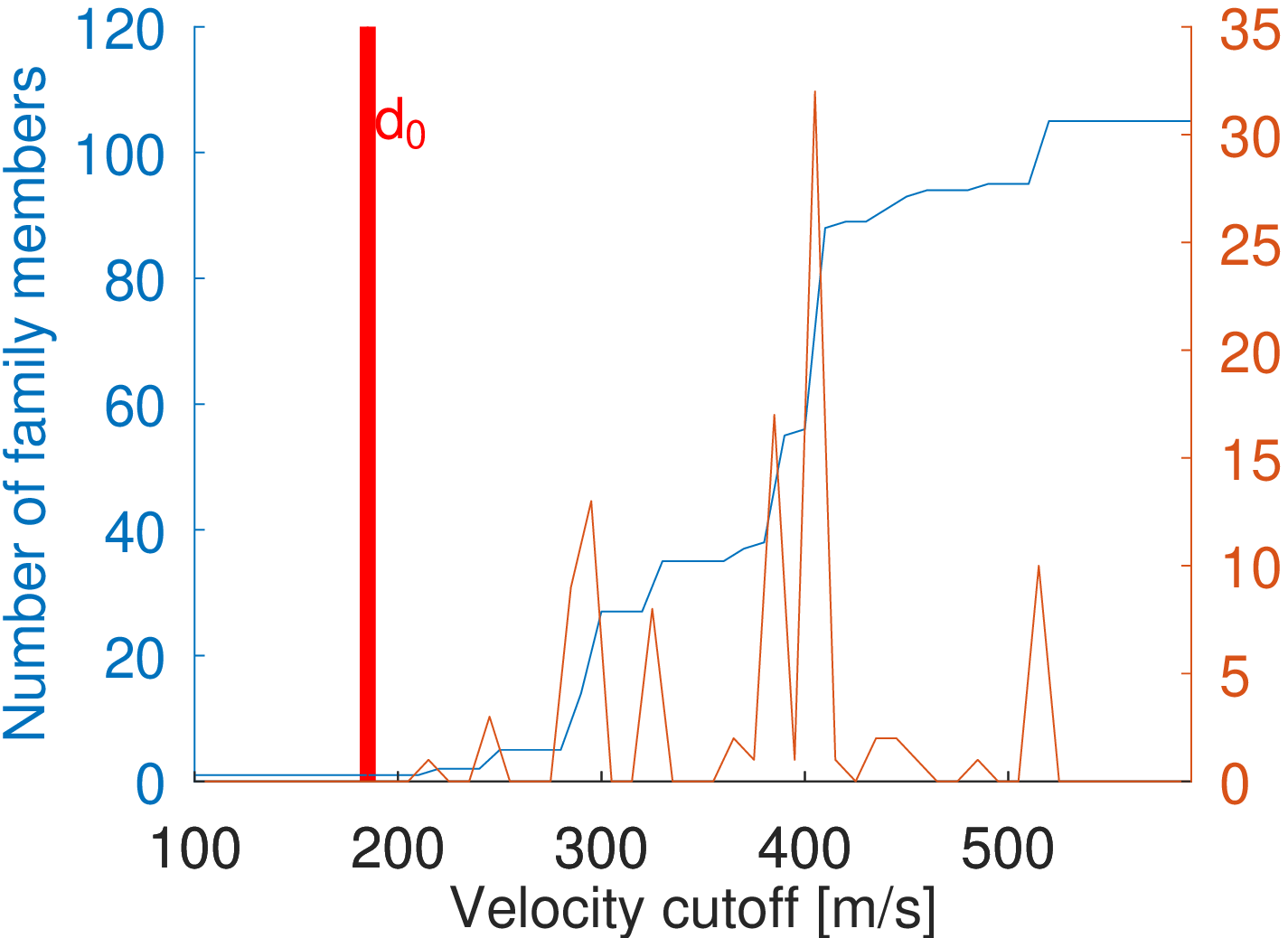}
  \end{minipage}%
  \begin{minipage}[c]{0.45\textwidth}
    \centering \includegraphics[width=2.3in]{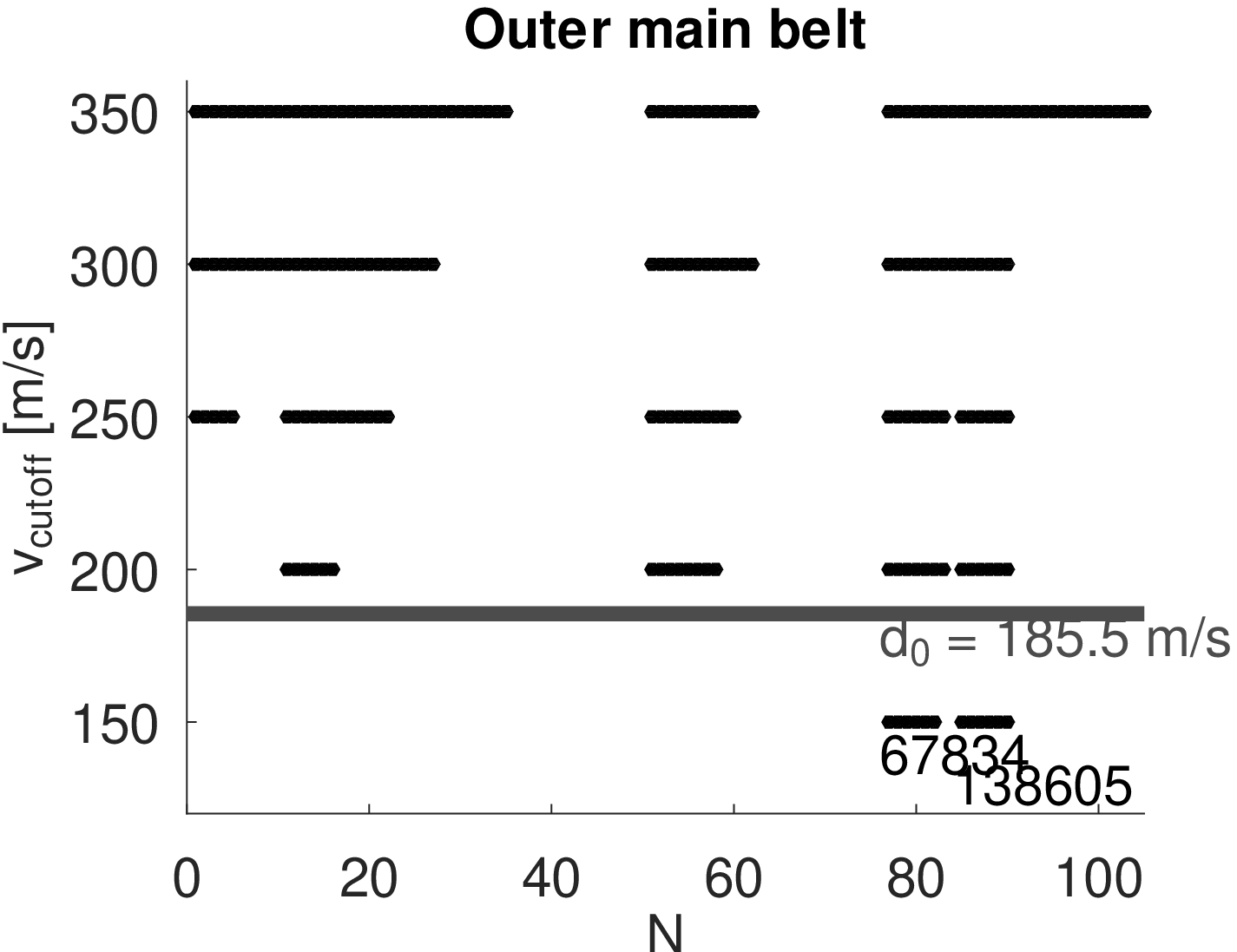}
  \end{minipage}
  \caption{The same as in figure (\ref{fig: groups_inner}), but for the outer
    main-belt.  The lowest numbered asteroid in the outer main-belt,
    3939 Huruhata is not associated with any known dynamical groups, which
    explains why no peak is observed at a cutoff equal to $d_0$.}
  \label{fig: groups_outer}
\end{figure*}

In the outer main-belt, the main family interacting with the ${\nu}_6$
resonance is that of Euphrosyne \citep{2014ApJ...792...46C}.
Figure~(\ref{fig: groups_outer}), left panel, shows how the family
members increased starting with the lowest numbered asteroid in the
inner part of the outer main-belt.  The peak observed at 410 m/s
occurs when there is the merging with the Euphrosyne family.
An analysis of the stalactite diagram for this region identifies
two groups: one around 67834 2000 VV53 and another around 138605 2000 QW177.
The first one is associated with the Euphrosyne family, while the
second is a completely new group, not associated with any currently
known asteroid family.

In the next section, we will analyze the physical and dynamical properties
of the groups identified in this work.

\section{Newly identified ${\nu}_6$ groups: properties}
\label{sec: nu6_groups}

As a first step, we attempt to extend the membership of identified
groups by including multi-opposition asteroids in ${\nu}_6$ antialigned
resonant states.  For this purpose, we use the approaches described
in \citet{2020MNRAS.496..540C} and \citet{2021CeMDA.133...24C}.
The groups identified in our previous analysis are used to train
a ML algorithm, whose hyper-parameters are first optimized through the
use of genetic algorithms \citep{chen_2004}, according to the procedure
described in \citet{2021CeMDA.133...24C}.  The best performing
algorithm is then used to select the population of multi-opposition
asteroids with the proper elements $(a,e,\sin{(i)})$ closest to
those of the training sample \citep{2020MNRAS.496..540C}.
These methods will identify asteroids whose proper orbital elements
are close to those of the identified families, but they will
not identify objects that could have been connected to the family using
traditional HCM.  Since the orbits of some multi-opposition asteroids
are known with some uncertainties, we believe that such a conservative
approach is reasonable.

The groups of interest will be then studied using standard HCM in a
domain of proper elements for all asteroids, resonant or not, to check
for what distance cutoff they connect to other local families.
This is important to check for families that are crossed
by the ${\nu}_6$ resonance and have a population of both resonant
and non-resonant members.  The limitations of this approach
have been discussed in section~(\ref{sec: nu6_sel}), see footnote (1).
Yet, it is a necessary step to see if the newly discovered families
extend beyond the limits of stable orbits in the ${\nu}_6$ resonance.

Finally, we looked to see if the taxonomic traits of resonant group members
were consistent with a common ancestor. We use the method of
\citet{2013Icar..226..723D} to obtain taxonomical information
for asteroids listed in the Sloan Digital Sky Survey-Moving Object Catalog data
(SDSS-MOC4; \citet{2001AJ....122.2749I}). We also look for bodies in the
WISE and NEOWISE, AKARI, or IRAS databases that have geometric albedo values
\citep{2012ApJ...759L...8M, 2011PASJ...63.1117U, 2010AJ....140..933R}. 
Under the assumption of an origin from a homogeneous parent body,
all members of a family should have compatible values of taxonomy and
albedo.

We start our analysis by looking at resonant groups in the inner main-belt.

\subsection{Inner main-belt}
\label{sec: inn_mb_g}

\begin{figure}
  \centering
  \centering \includegraphics[width=3.0in]{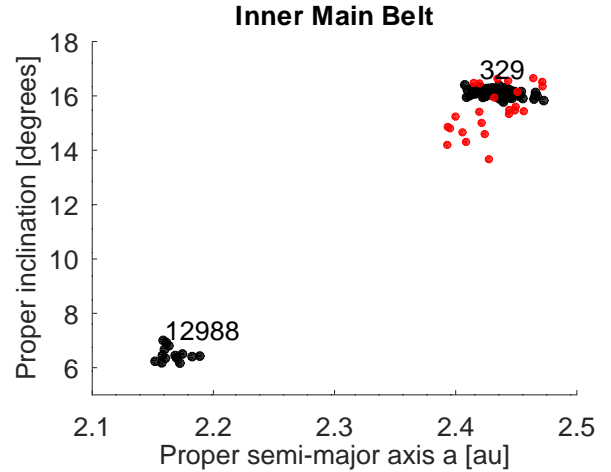}
  \caption{The $(a,\sin{(i)})$ distribution of the identified numbered
    (black full circles) and multi-opposition (red full circles)
    members of resonant groups in the inner main-belt.}
  \label{fig: phys_prop_inner}
\end{figure}

Two main groups were identified in the inner main-belt, that of 34804,
associated with the Svea asteroid family, and the new Tiffanykapler
group at lower values of $a$.  We first extended these two groups
in the domain of resonant multi-opposition asteroids.  The best performing
algorithms after five generations were an Extra Tree and a Decision Tree
classifier.  For the Extra Tree classifies, we had these hyper-parameters:
bootstrap=False, criterion="gini", max$\_$features=0.7,
min$\_$samples$\_$leaf=13, min$\_$samples$\_$split=17, n$\_$estimators=100.
For the case of the Decision Tree estimator, we had:
max$\_$depth=7, min$\_$samples$\_$leaf=14, min$\_$samples$\_$split=19.
Interested readers can find more information on the meaning of the
hyper-parameters name and their effect on the Scikit-learn documentation
page (https://scikit-learn.org/stable/, \citet{scikit-learn}), or
in \citet{2021CeMDA.133...24C}. 24 new possible members of the Svea
family were found by these methods, while no multi-opposition
asteroid in our database appears to be a possible member of the
Tiffanykapler group (see figure~(\ref{fig: phys_prop_inner})).

The HCM analysis of these two groups shows that the Svea resonant group
connects with a nearby family for a cutoff of 150 m/s, while the Tiffanykapler
joins the Flora family at a distance of 110 m/s.  Both groups are fairly
isolated and identifiable in proper elements domains.
Finally, no SDSS taxonomical information is available for any members
of the two resonant groups.  Albedo information is available 
only for 13 members of the Svea resonant group and for 1
of the Tiffanykapler.  The albedo of 12 members of the Svea group
is below 0.08, and compatible with that of the family, which belongs
to the C-complex.  One object, 59626 1999 JA75 has an albedo of 0.187
and could be an interloper.  The only object with an albedo value
in  Tiffanykapler is 65010 2002 AR82, with $p_V = 0.457$.  The largest
object in this group is 12988 Tiffanykapler itself, with an
estimated diameter of 1.45 km, using this albedo value and the
asteroid absolute magnitude of 15.82.  Other group members are all smaller
than 1.0 km in diameter.

\subsection{Central main-belt}
\label{sec: cen_mb_g}

\begin{figure}
  \centering
  \centering \includegraphics[width=3.0in]{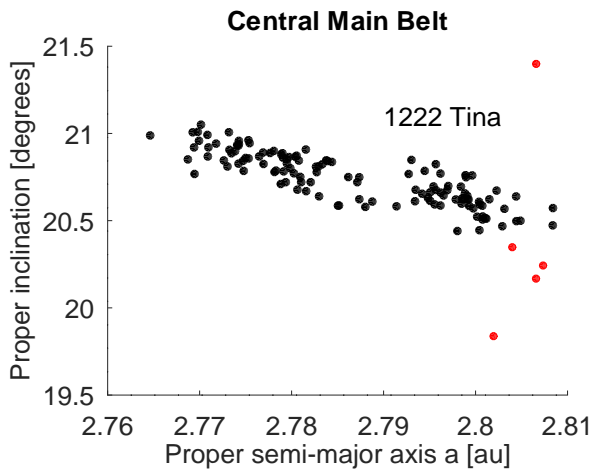}
  \caption{The $(a,\sin{(i)})$ distribution of the identified numbered
    (black full circles) and multi-opposition (red full circles)
    members of the Tina asteroid family in the central main-belt.}
  \label{fig: phys_prop_central}
\end{figure}

The main group of asteroids in the antialigned configuration in the central
main-belt is the Tina family \citep{2011MNRAS.412.2040C}.  The best performing
algorithm to extend this family in the domain of multi-opposition
asteroids was a Decision Tree classifier with max$\_$depth=10,
min$\_$samples$\_$leaf=4, and min$\_$samples$\_$split=5.  Five
multi-opposition asteroids
were found to be likely members of the Tina family.  Their orbital
location is shown in figure~(\ref{fig: phys_prop_central}).
 
Because of its orbital location, entirely located in a stable island
in the unstable orbital region caused by the ${\nu}_6$ secular resonance,
the Tina family is rather isolated and quite distinguishable from the
surrounding families.  All the asteroids in the stable island are connected
to the family by HCM for a cutoff distance of 130 m/s, and the group
merges with nearby families only at a distance of 280 m/s.  A population
of 44 asteroids at higher eccentricities than those of the family
(see figure~(\ref{fig: nu6maps}), center-right panel) has been
identified in this work for the first time. Its dynamical evolution
and possible origin from the Tina family is an interesting topic for
future research.

\begin{figure}
  \centering
  \centering \includegraphics[width=3.0in]{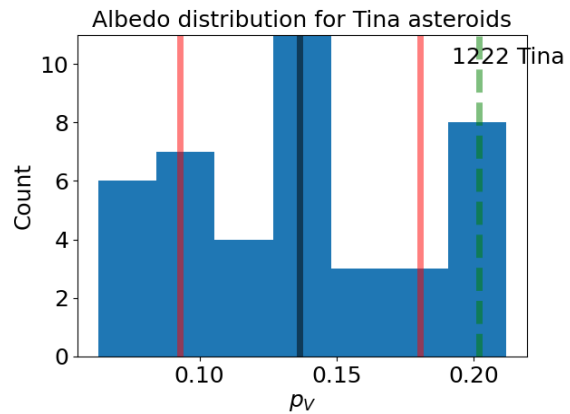}
  \caption{An histogram of albedo values in the Tina family.  The black
    vertical line show the mean albedo value, while the vertical red lines
    shows the limits of the albedo range, as defined in the text.  The green
    vertical dashed line displays the albedo of 1222 Tina.}
  \label{fig: alb_tina}
\end{figure}

SDSS taxonomy is available for only one object, 81986 2000 QQ123,
whose X-type is compatible with the taxonomy of 1222 Tina.  42 asteroids
have values of albedo, mostly provided by the NEOWISE database.
The albedo distribution of these asteroids is rather difficult
to interpret.  The range covered by the mean value, plus
or minus one standard deviation, is 0.137$\pm$0.044.  However,
the albedo of 1222 Tina, 0.202 itself is outside this range, and there
are six asteroids with albedos lower than the values in the range.
This problem was already noticed in \citet{2011MNRAS.412.2040C},
and the new data confirmed this trend.  Figure~(\ref{fig: alb_tina})
displays a histogram of the albedo values.  The unusual albedo values,
generally quite large for an X-type asteroid family, and their unusual
distribution will be investigated later on in this paper.

\subsection{Outer main-belt}
\label{sec: out_mb_g}

\begin{figure}
  \centering
  \centering \includegraphics[width=3.0in]{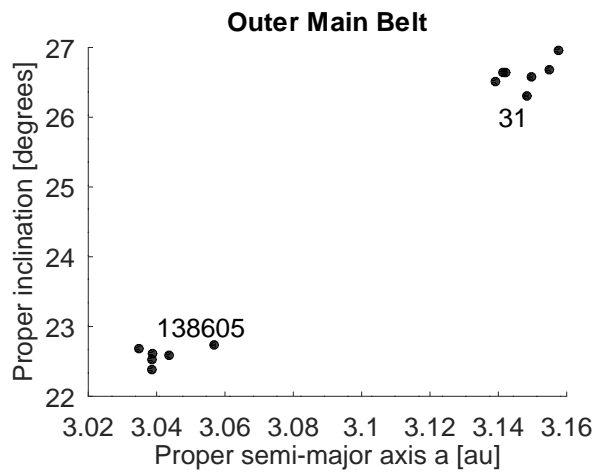}
  \caption{The $(a,\sin{(i)})$ distribution of the identified numbered
    members (black full circles) of resonant groups in the outer main-belt.}
  \label{fig: phys_prop_outer}
\end{figure}

Two groups have been identified in the outer main-belt, those of 67834
2000 VV53, associated with the Euphrosyne family, and the new
group of 138605 QW177.  The best-performing algorithms for
connecting these groups to the multi-opposition population
were an Extra Tree Classifier with the same hyper-parameters as
in section~(\ref{sec: inn_mb_g}) and a Gaussian Naive Bayes with no
optimized hyper-parameter.  Neither identified new members of the
two groups among the 24 multi-opposition asteroids in the region.
The $(a,\sin{(i)})$ orbital distribution of the two groups
is displayed in figure~(\ref{fig: phys_prop_outer}).

The HCM analysis of these two groups shows that the first one connects
to the surrounding Euphrosyne family already at a cutoff of 30 m/s,
while the other group is rather isolated, only connecting to nearby
families for the rather large value of the cutoff of 130 m/s.
No SDSS information is available for any members of both groups.
NEOWISE albedo information is available for 2 members of the Euphrosyne
and 4 members of the 138605 group.   All six asteroids have albedos
lower than 0.08, compatible with a C-complex taxonomy, which is
consistent with the case of the Euphrosyne family.
The largest member of the 138605 group is 417866 2007 LY33,
with an estimated diameter of 6.077 km.  The remaining asteroids
in the group have estimated diameters lower than 5.2 km.

\section{Dating the new resonant families}
\label{sec: dat}

Having identified two new possible asteroid groups, here we attempt to
infer information about their ages.  Both the 12988 and 138605 are
fairly small groups, with 14 and 6 members each.  Dating methods
using the slope of the V-shape of family members in the $(a,1/D)$ domain
\citep{2015Icar..257..275S}, or the Yarko-Yorp approach of
\citet{2006Icar..182..118V}, among others, will not
work for such groups, because of small number statistics issues.  However, both
groups are ideally suited for being studied with methods
designed for young clusters, like the Backward Integration Method
(BIM, \citet{2002Natur.417..720N}) and the Close Encounters Method
(CEM, \citet{2010Natur.466.1085P})

In the BIM method, from time-reversal numerical simulations, past
discrepancies in the longitudes of the pericenter $\varpi$ and node
$\Omega$ of family members with respect to those of the claimed parent body
are obtained.  These differences should converge to values approaching
zero at the time of family formation. For asteroids inside the
${\nu}_6$ secular resonance, where $\varpi$ is in a locked
state with the longitude of pericenter of Saturn, the
method can only use convergence in $\Omega$, but the principle
remains the same.

The CEM approach works by integrating into the past multiple clones of
the parent body and of the other family members. Close encounters that occur
at low relative distances and speeds between
two clones are recorded, and the median value of these times is used to
estimate the asteroid pair's age. The cutoff values for
relative distances and speed of the encounters are defined in terms of
the Hill's radius and escape velocity of the primary body.
More technical details on both these methods, as implemented by our group,
can be found in the methods section of \citet{2020NatAs...4...83C}
\footnote{As discussed in \citet{2020NatAs...4...83C}, fission
pairs may occur inside relatively young asteroid families.  We looked
for candidate pairs, characterized by relative distance in proper
elements, as computed with equation (4) in \citet{2020NatAs...4...83C}, of less
than 5 m/s, and a mass ratio (see equation (5) in \citet{2020NatAs...4...83C})
of less than 0.3, in both the 12988 and 138605 groups.  No candidate pair
was identified in either families.}.

\begin{figure*}
  \centering
    \begin{minipage}[c]{0.45\textwidth}
    \centering \includegraphics[width=2.3in]{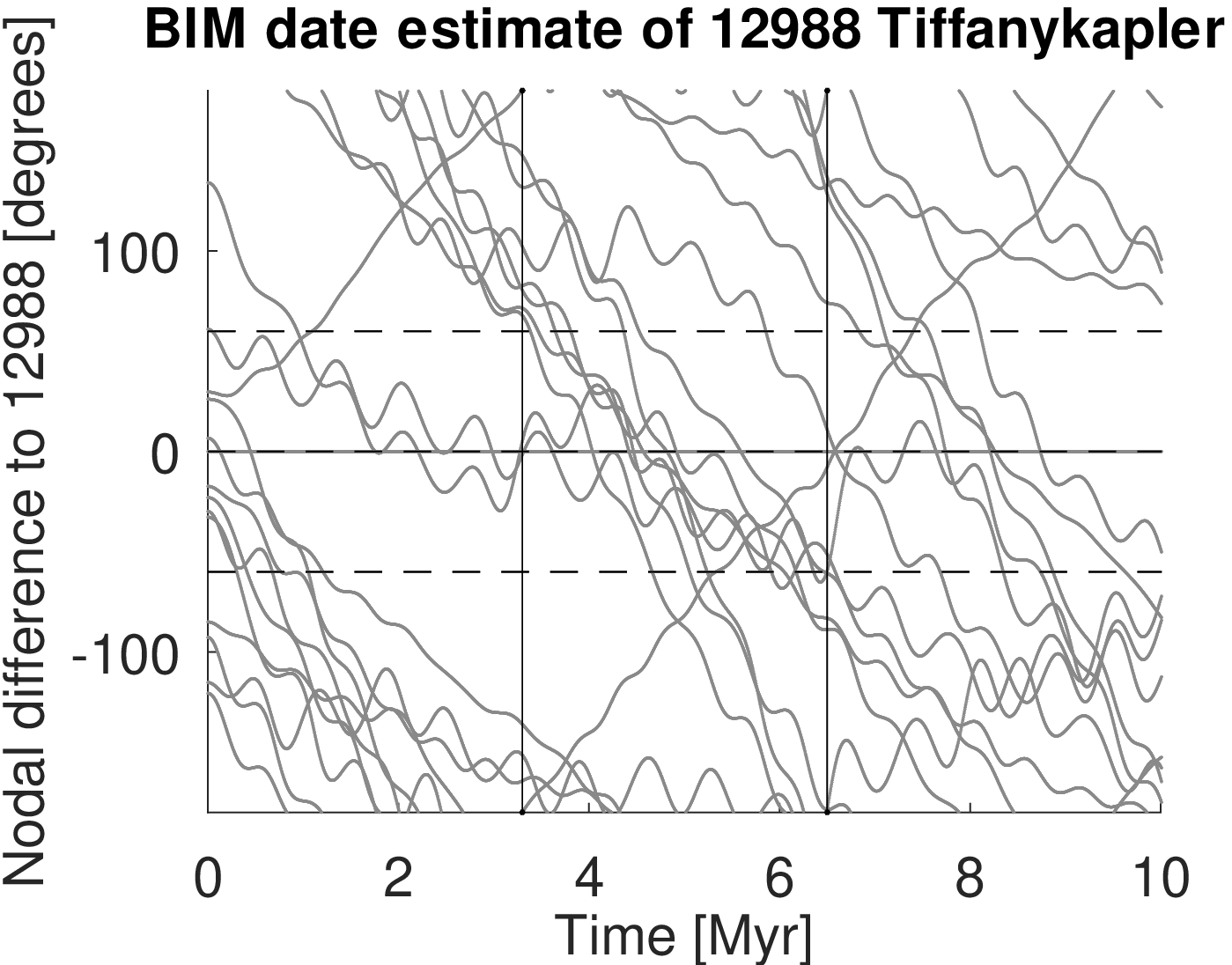}
  \end{minipage}%
  \begin{minipage}[c]{0.45\textwidth}
    \centering \includegraphics[width=2.3in]{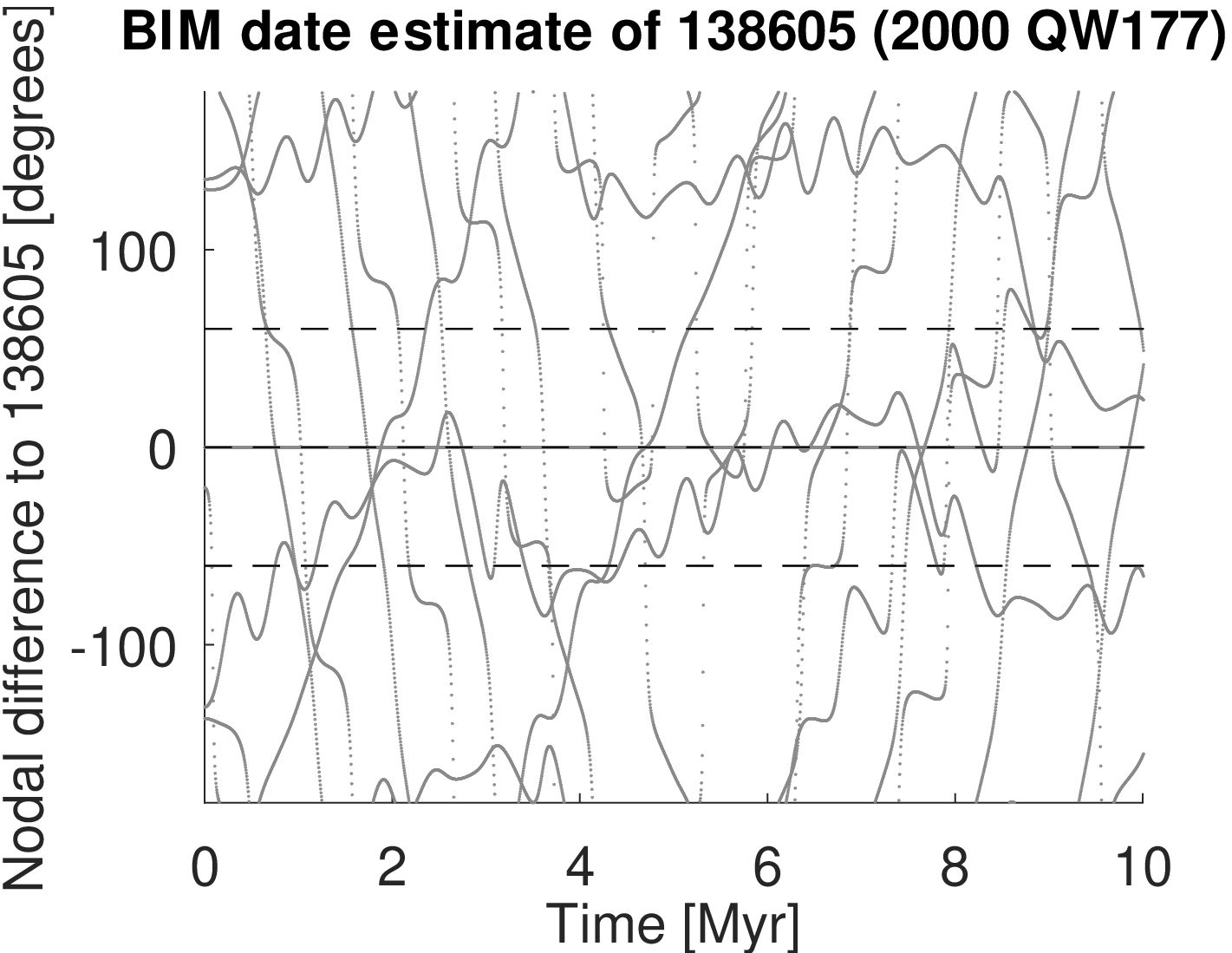}
  \end{minipage}

  \caption{Past convergence of the differences in $\Omega$ with respect
    to the alleged parent body for members of the 12988 Tiffanykapler
    (left panel) and 138605 QW177 (right panel) groups.}
\label{fig: BIM_groups}
\end{figure*}

Figure~(\ref{fig: BIM_groups}) displays the past convergence of differences
in $\Omega$ with respect to the possible parent body of members of the
12988 Tiffanykapler (left panel) and 138605 QW177 (right panel) groups.
Since we are neglecting in this simulation the effect of non-gravitational
force, convergence should occur to within $\pm 60^{\circ}$.We show
these levels as horizontal dashed lines in the figure.  We find a
possible solution for the 12988 group in the time range from 3.3 to 6.5 My
ago (see figure~(\ref{fig: BIM_groups}, left panel, vertical dashed lines)).
No solution was found for the 138605 group.  According to these
results, 12988 Tiffanykapler could be the first young asteroid family
to ever be found in a linear secular resonance state.

\begin{figure}
  \centering
  \centering \includegraphics[width=3.0in]{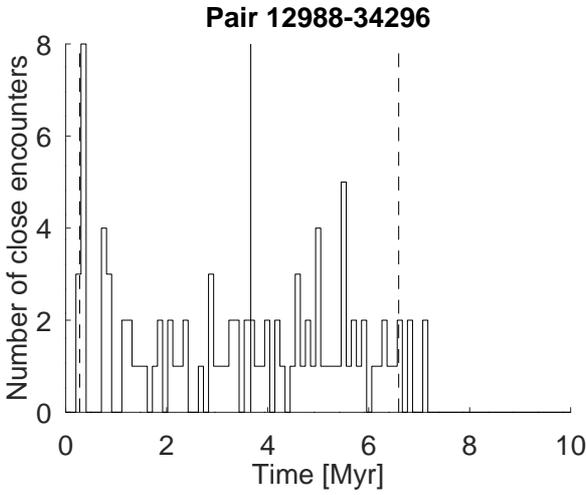}
  \caption{An example of statistics of close encounters between 12988
    and 34296, a member of the Tiffanykapler group.  The vertical line
    shows the median value of the age estimate, while the vertical dashed lines
    display the 5th and 95th percentiles of the distribution, used to estimate
    the error on the pair age.}
\label{Fig: CEM_Example}
\end{figure}

To confirm this exciting hypothesis, we turn our attention to the results from
the CEM approach.  Figure~(\ref{Fig: CEM_Example}) shows an example
of how statistics of close encounters in the past can be used to estimate
the age of an asteroid pair.  The median value of the distribution is
used to obtain the age estimate, while the errors are obtained by considering
the 5th and 95th percentiles.  For the pair shown in the figure the age
was of $3.665^{+2.929}_{-3.388}$, and there were 95 encounters that satisfied
our selection criteria in this simulation.  While all the asteroid
pairs in the Tiffanykapler groups have significant statistics of close
encounters, all with ages compatible with the range found from BIM,
most asteroid pairs in the 138605 groups have less than 10 encounters
over the 10 My length of the simulation.   Both results from BIM and
CEM suggest that either the 138605 is older than 10 My, or that it
may be an artifact of the HCM procedure described in
section~(\ref{sec: nu6_groups}).   

\begin{table}
  \begin{center}
    \caption{Age estimates for pairs in the 12988 Tiffanykapler cluster
      obtained with CEM.}
    \label{Table: 12988_CEM}
    \begin{tabular}{|c|c|c|}
      \hline
      Asteroid  & Number of & Age   \\
      Id.       & enc.      & [My] \\
      \hline
      34296  & 95 & $3.665^{+2.929}_{-3.388}$ \\
      65010  &124 & $3.273^{+3.519}_{-3.060}$ \\
      112247 & 68 & $3.540^{+3.260}_{-2.685}$ \\
      131528 & 98 & $3.163^{+3.521}_{-2.547}$ \\
      149823 & 85 & $2.893^{+3.461}_{-2.547}$ \\
      227780 & 82 & $2.993^{+3.433}_{-2.830}$ \\
      340833 & 88 & $3.788^{+2.788}_{-3.618}$ \\
      380442 & 89 & $2.614^{+3.157}_{-2.377}$ \\
      389850 & 83 & $3.192^{+3.396}_{-3.020}$ \\
      408140 & 45 & $3.829^{+2.919}_{-3.571}$ \\
      464017 & 73 & $2.491^{+4.095}_{-2.465}$ \\
      483384 & 92 & $2.380^{+3.735}_{-2.005}$ \\
      495612 & 55 & $4.132^{+2.855}_{-3.976}$ \\
\hline
\end{tabular}
\end{center}
\end{table}

Our results for the pairs in the 12988 Tiffanykapler group are summarized
in Table~(\ref{Table: 12988_CEM}).  If we take as an estimate of the error
the larger value between the upper and lower age estimate range, which,
for the first asteroid pair would mean taking 3.388 My as an
estimate of the standard deviation ${\sigma}_1$ for the age $x_1$, then
the asteroid age can be estimated using the weighted mean formula:

\begin{equation}
\overline{x}=\frac{\sum_{i=1}^{n_{pairs}}\left(\frac{x_i}{{{\sigma}_i}²}\right)}{\sum_{i=1}^{n_{pairs}}\frac{1}{{{\sigma}_i}^2}},
\label{eq: weighted_mean}
\end{equation}

\noindent where $n_{pairs}$ is the number of asteroid pairs, in our case 13.
The standard error of the weighted mean is then given by:

\begin{equation}
  {\sigma}_{\overline{x}}=\sqrt{\frac{1}{\sum_{i=1}^{n_{pairs}}\frac{1}{{{\sigma}_i}^2}}}.
    \label{eq: std_w_m}
\end{equation}

\noindent Using the data from table~(\ref{Table: 12988_CEM}) we obtain an
estimated age for the 12988 Tiffanykapler family of $3.05\pm 1.15$~My.
This makes the Tiffanykapler the first ever identified young asteroid
family inside a linear secular resonance.  In the next section, we will
investigate what information can be obtained from the resonant configurations
of these two groups.

\section{Secular constraints on the initial ejection velocity field of
  the 12988 Tiffanykapler family}
\label{sec: cons_quant}

\begin{figure*}
  \centering
    \begin{minipage}[c]{0.45\textwidth}  
    \centering \includegraphics[width=2.3in]{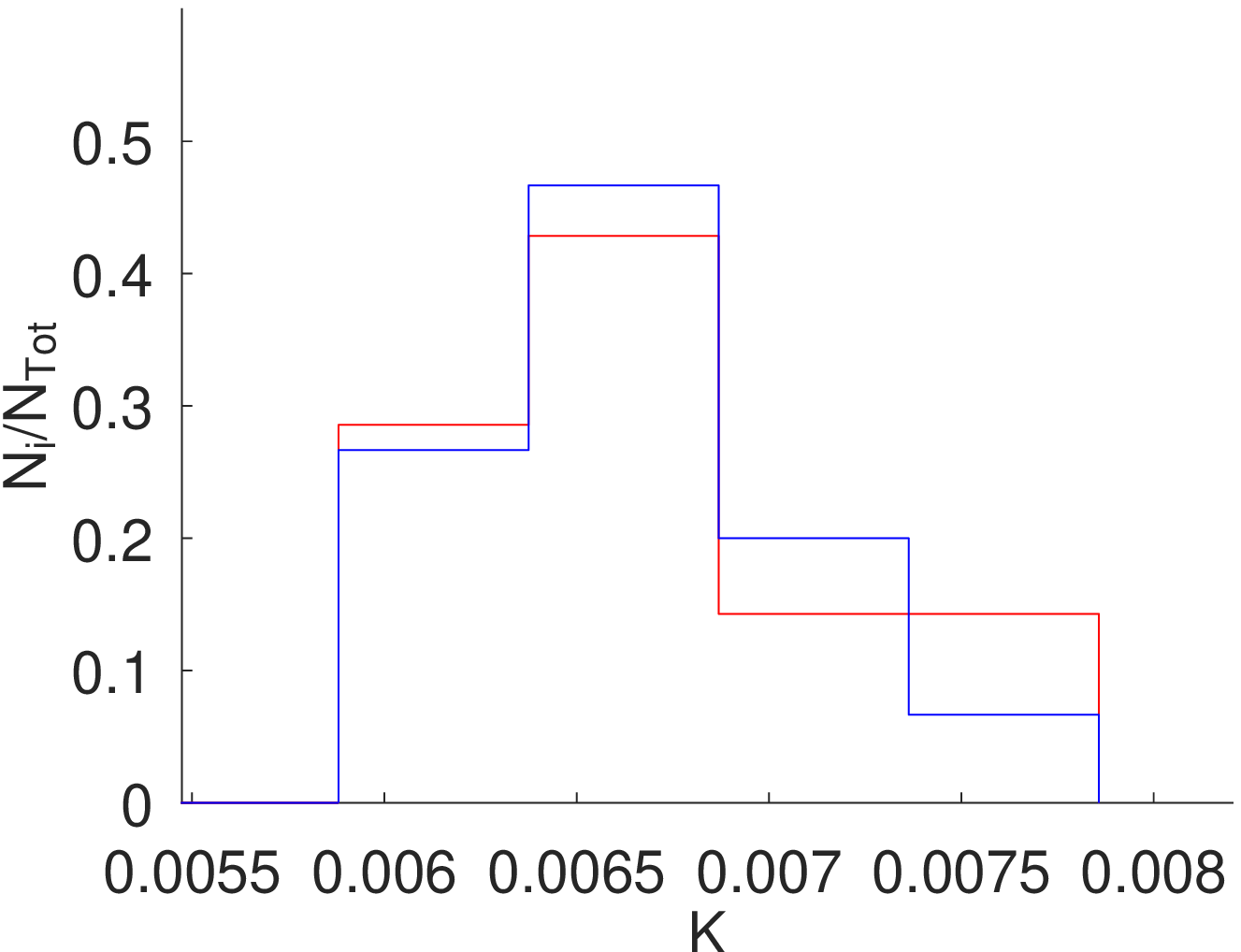}
  \end{minipage}%
  \begin{minipage}[c]{0.45\textwidth}
    \centering \includegraphics[width=2.3in]{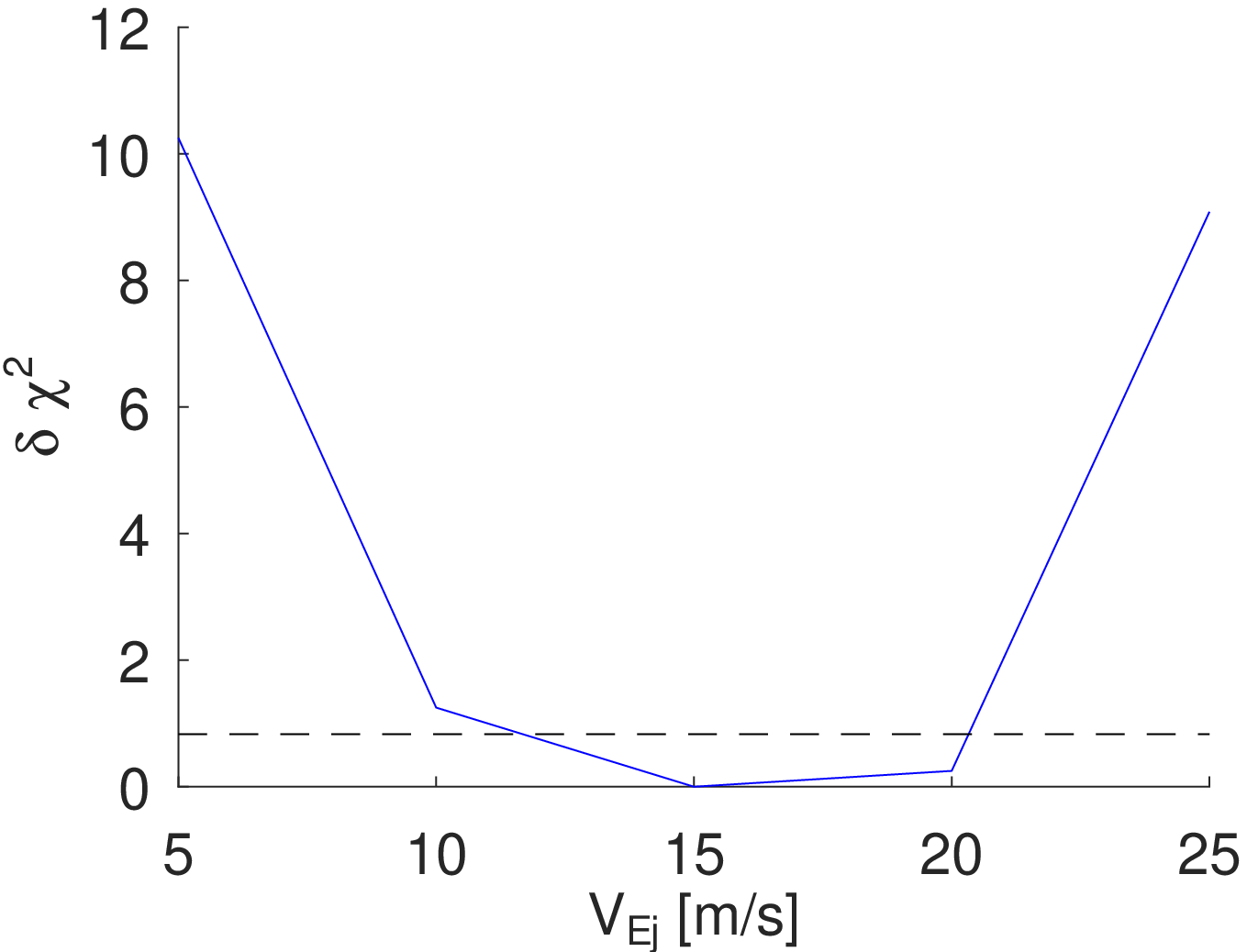}
  \end{minipage}

  \caption{Left panel: a normalized histogram of the distribution
    of $K_2^{'}$ values for members of the Tiffanykapler family (red line)
    and of the simulated asteroid family that best fit the data (blue line).
    Right panel:  Dependence of $\Delta {\chi}^2$ as a function of the
    $V_{EJ}$ parameter.  The 97.5\% confidence level limit is
    identified by the horizontal dashed line.}
\label{fig: k2_cons}
\end{figure*}

In this section we will investigate what constraints can be obtained
on the ejection velocity field of the Tiffanykapler family from
its secular nature.  Since there are only six
members in the 138605 QW177 group, these statistical methods cannot
be applied to this family.  
As discussed in \citet{2011MNRAS.412.2040C}, at the simplest level of
perturbation theory the quantity:

\begin{equation}
  K_2^{'}=\sqrt{1-e^2}[1-\cos{(i)}],
\label{eq: k2}
\end{equation}

\noindent is preserved for asteroids in the ${\nu}_6$ resonance in the
conservative case.  Simulations performed by \citet{2011MNRAS.412.2040C}
showed that this quantity is also preserved when non-gravitational forces are
considered.  Current values of the $K_2^{'}$ quantities for
members of the Tiffanykapler family are displayed in the left
panel of figure~(\ref{fig: k2_cons}).

The preservation of the original values of $K_2^{'}$ permits to estimate
the initial ejection velocity field with a method not generally available
for non-resonant families (\citet{2006Icar..183..349V}, see also
the review in \citet{2018P&SS..157...72C}).
First, we assume that asteroids are ejected with an isotropic ejection
velocity field, and that the standard deviations $V_{SD}$ are proportional to
the asteroid size through the parameter $V_{EJ}$, via the relationship:

\begin{equation}
  V_{SD}=V_{EJ}\cdot(\frac{5~km}{D}),
  \label{eq: V_ej}
\end{equation}

\noindent where $D$ is the asteroid diameter in km.  While this method
may not provide an accurate measure of the ejection velocities
for asymmetric fields, it can provide useful constraints on its
magnitude.  We create multiple synthetic asteroid families with various
$V_{EJ}$ parameter values, calculate their $K_2^{'}$ quantities, and
determine which of the simulated families is best compatible with the
currently observed $K_2^{'}$ distribution.
We can utilize a ${\chi}^2$-like variable defined as equation~(\ref{eq: chi2})
to quantitatively test this: 

\begin{equation}
  {\chi}^2=\sum_{i=1}^{N_{int}} \frac{(q_i-p_i)^2}{q_i}.
    \label{eq: chi2}
\end{equation}

\noindent $N_{int} = 6 $ is the number of intervals used to obtain the
histogram of the $K_2^{'}$ distribution. $q_i$ and $p_i$ are the
numbers of real and simulated objects in the $i-th$ interval, respectively.
Once values of $\Delta {\chi}^2 = {\chi}^2-{{{\chi}^2}_{min}}$ are obtained,
where ${{{\chi}^2}_{min}}$ is the minimum value of ${\chi}^2$, the ${\chi}^2$
probability distribution function can be used to determine the 97.5\%
confidence level (see \citet{2018P&SS..157...72C} for a more in-depth
description of this method). Our results are shown in the right panel
of figure~(\ref{fig: k2_cons}).  Our analysis yields a value
of $V_{EJ} = 15^{+6}_{-3}$ m/s.  

\section{Tina family: albedo distribution and high eccentricity population}
\label{sec: Tina_extra}

Two open questions were raised by our investigation of the Tina family:
the unusual albedo distribution of Tina asteroids (see
figure~(\ref{fig: alb_tina})) and the origin of the high eccentricity
population (see figure~(\ref{fig: nu6maps}), panel 4).

\begin{figure*}
  \centering
  \begin{minipage}[c]{0.45\textwidth}
    \centering \includegraphics[width=2.3in]{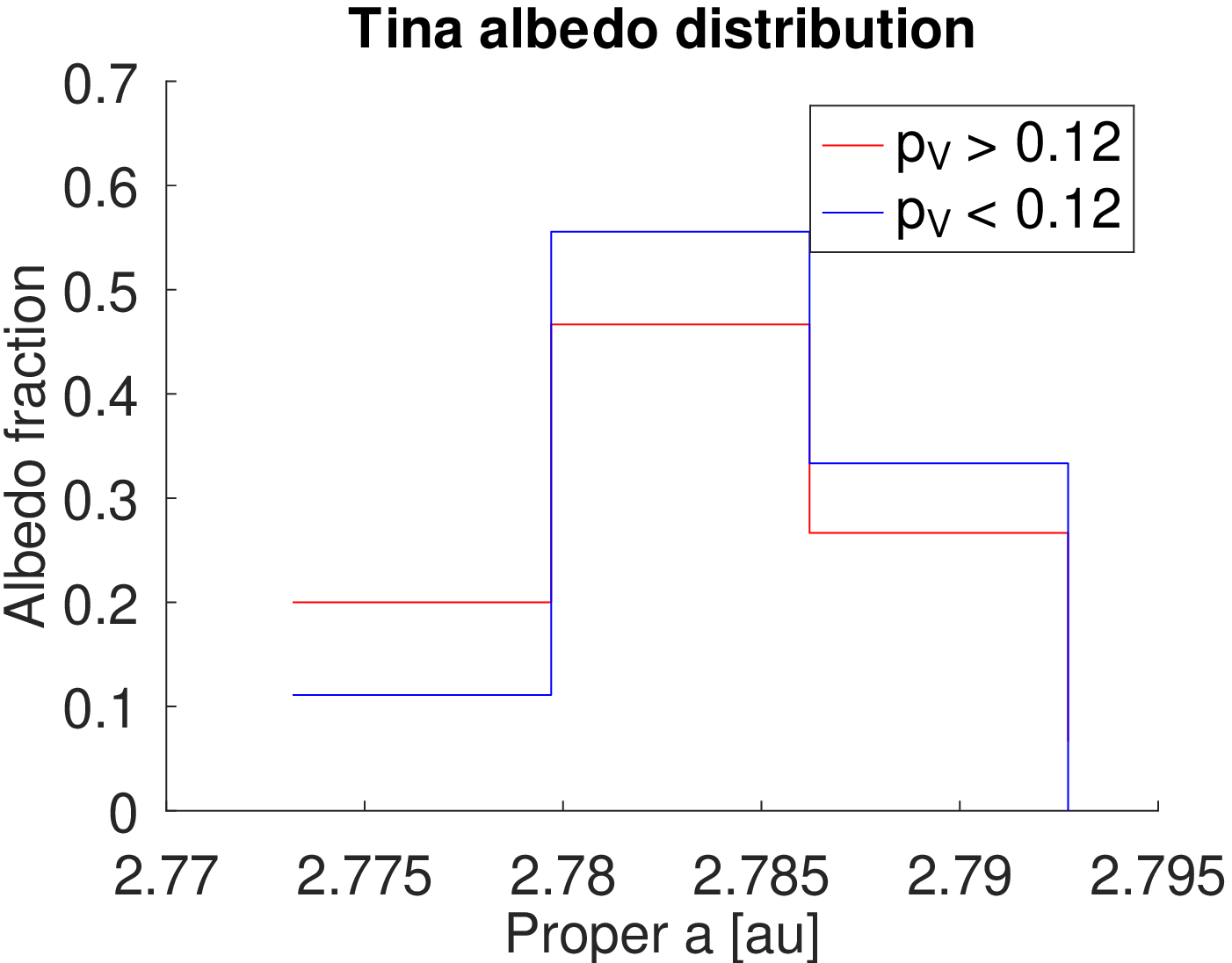}
  \end{minipage}%
  \begin{minipage}[c]{0.45\textwidth}
    \centering \includegraphics[width=2.3in]{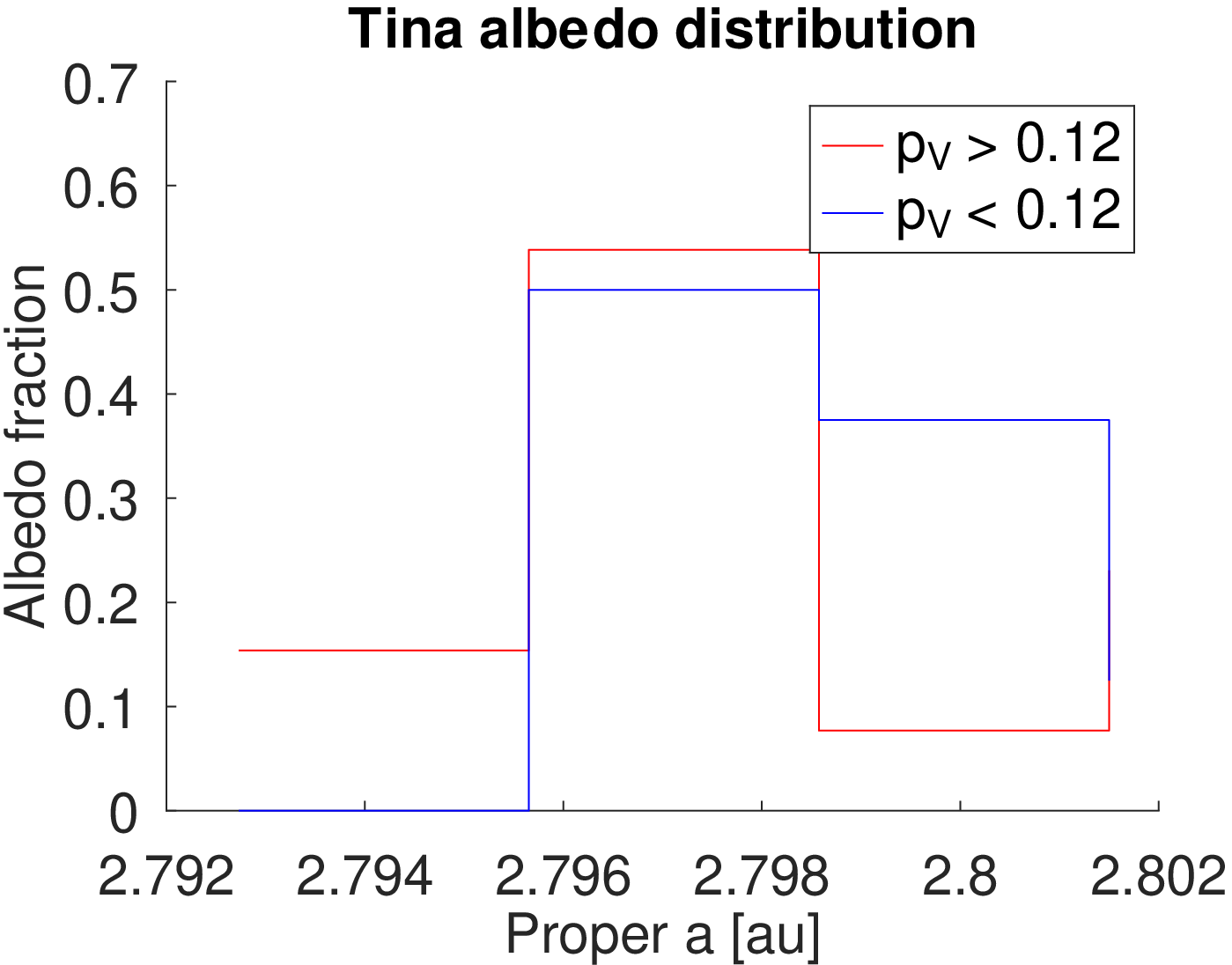}
  \end{minipage}
  \caption{The percental distribution of low (blue curve, $p_V < 0.12$) and
    high (red curve, $p_V > 0.12$) albedo asteroids in the left and right sides
    of the Tina family V-shape.}
\label{fig: albedo_distr}
\end{figure*}

To answer the first question, we divided the sample of objects with albedo
information in two parts, those with values of $a$ less than that of
1222 Tina, and those with larger values.  Since the family center
of mass is essentially located at the orbit of Tina, the first population
corresponds to the left side of the family V-shape in the
$(a, 1/D)$ domain, as defined in \citet{2015Icar..257..275S}, while the
second belongs to the right side.
Then, we further divide asteroids according to their albedo, with
a first sample with $p_V < 0.12$, usually associated to a C-complex
taxonomy, and a second sample with $p_V > 0.12$, usually related to 
S-complex compositions.  The percental distribution of these
four groups are shown in figure~(\ref{fig: albedo_distr}).

The distributions of low- and high albedo objects on the left and right
side of the Tina family are highly compatible.  A two-sample Kolmogorov-Smirnoff
test (or K-S test, \citep{smirnov1939estimate}) dismissed the probability
of the two distributions being different to the null hypothesis level of 5\%.
This result suggests that a common origin should be found for
both populations.  The Tina family may be characterized by an unusually
broad distribution of albedo values, not commonly observed in other
parts of the main-belt for X-type families.

Concerning the high-e Tina asteroidal population, as discussed
in section~(\ref{sec: nu6_sel}), resonant proper elements are more
appropriate for the case of resonant asteroids.  Two approaches
for obtaining resonant elements were discussed in \citet{2011MNRAS.412.2040C},
one based on Fourier analysis of the equinoctial elements
$(e\cos(\varpi -{\varpi}_6), e \sin(\varpi -{\varpi}_6))$, where
the second frequency of largest amplitude is identified as
the resonant proper frequency $g_{\sigma}$, and its associated
amplitude is the resonant eccentricity $e$. The second method
is based on the conservation of the $K_2^{'}$ quantity and on
taking as proper $e$ the value of $e$ when $\sigma =180^{\circ}$
and $\frac{d \sigma}{dt} > 0$.  As in \citet{2011MNRAS.412.2040C},
here we will use the first approach, since the second method
does not provide values of the proper frequency $g_{\sigma}$.

\begin{figure}
  \centering
  \centering \includegraphics[width=3.0in]{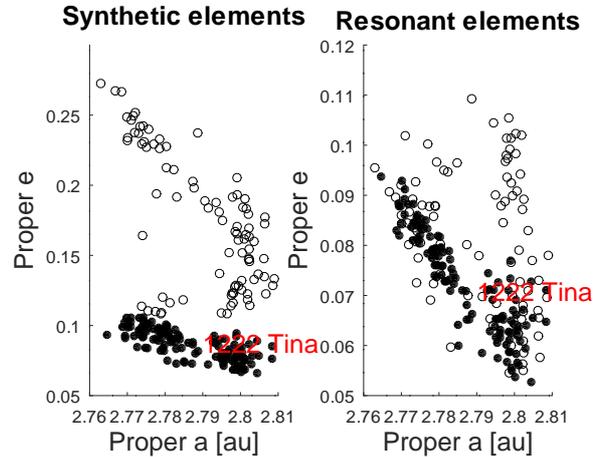}
  \caption{On the left panel we display the projection in the synthetic
    proper $(a,e)$ domain of Tina family members, as obtained with HCM,
    and other anti-aligned asteroids at higher eccentricity in the same
    stable island.  Tina family members are shown as full black dots,
    while the other asteroids are shown as black circles.  The right
    panel displays the position of the same asteroids in the synthetic
    proper element $(a,e)$ domain.  The position of 1222 Tina itself
    is identified by the red label.}
\label{Fig: Tina_res_el}
\end{figure}

Our results for both synthetic and resonant elements are shown
in figure~(\ref{Fig: Tina_res_el}).  Quite interestingly, values of
resonant proper $e$ are significantly smaller than those of
the synthetic $e$.  The bulk of the high-$e$ population is
still not connected to the Tina dynamical family even if we used
HCM in the new domain of resonant proper elements. But, it connects
to the family at a higher cutoff of 100 m/s.  This behavior is consistent
with what is observed for other families halos.  The concept
of an asteroid family halo was first introduced by \citet{2013Icar..223..844B}
and refers to an extended population of asteroids that is not recognized
as part of a family by standard HCM approaches, but it is
quite likely to have been produced by the collision event that produced
the dynamical family.  Experience with other families halos
shows that many objects in these population may connect to the family
core when higher values of cutoff are considered \citep{2013MNRAS.433.2075C}.
This high-$e$ population of Tina-like asteroids may be the first
example of a ``resonant halo'' ever to be identified. The main difference
with respect to the halo of non-resonant families being that the dimensions
of this halo are limited by the available area in the stable island. One object,
599705 (2010 TD202), is very close to the resonant separatrix and becomes
unstable on timescales of a few Kyr when time-reversal simulations are
carried out. This can be an example of observable dynamical erosion of an
asteroid family in action, as defined in \citet{2010MNRAS.403.1834C}.

\section{Conclusions}
\label{sec: concl}

The main goal of this paper was to update our knowledge on the
population of asteroids in stable states {\bf inside} the ${\nu}_6$
secular resonance, and to identify new asteroid groups among this
population.  As discussed in \citet{2018P&SS..157...72C},
asteroid groups interacting with secular resonances are important
for the information they can provide on the orbital evolution
caused by Yarkovsky effect and for the independent constraints
on the age and ejection velocity fields that can be inferred from their
dynamical configurations.

Using a barrage of methods involving artificial neural
networks ($ANN$), clustering methods, supervised machine learning
approaches optimized through the use of genetic algorithms, we
were able to obtain a sample of 15 asteroids on aligned orbits
and 1669 asteroids on anti-aligned orbits, the largest database
so far for this asteroidal population.  We obtain the largest
sample of resonant arguments images ever produced, with 10004 images,
which may be quite valuable for future studies of automatic
$ANN$ recognitions of asteroid resonant behavior \citep{2021MNRAS.504..692C}.
We retrieved the three
known families crossed by the ${\nu}_6$ resonance: the Tina, Euphrosyne,
and Svea groups.  Two new asteroid groups were identified for
the first time in this work: those of Tiffanykapler in the
inner main-belt and of 138605 QW177 in the outer belt.  Both
are completely made of anti-aligned asteroids, which makes them
only the second and third groups to have this property, after the
Tina family.  The Tiffanykapler group is a young asteroid family with
an age of $3.0\pm1.2$ My, confirmed by two independent dating methods
based on time-reversal numerical integration, the Backward Integration
Method (BIM) and the Close Encounters Method (CEM).  The unique resonant
configuration of the Tiffanykapler family, and conserved quantities of
the ${\nu}_6$ secular resonance allow setting constraints on the family
ejection velocity field parameter, with a value of $V_{EJ} = 15^{+6}_{-3}$
m/s.  This is the first-ever identified young family in the ${\nu}_6$
secular resonance, and only the second group, after the Zelima sub-family
\citep{2019MNRAS.482.2612T}, for which accurate estimates on the family
age and original ejection velocity field can both be obtained.  As more numerous
databases of asteroid proper elements are produced, we expect that more
and more young asteroid families will be encountered, as recently
observed for the Zelima and Tiffanykapler groups.

A new sample of high-$e$ asteroids near the Tina dynamical family
was also observed in this work. These objects become connected to the rest of
the Tina family at a higher velocity cutoff when using HCM in the
domain of resonant elements.  This behavior, and the comparable
values of albedos of these objects with the rest of the Tina family
suggest that they could belong to the first ``resonant halo''
ever observed in the main-belt.  They would be part of an extended
Tina family not detectable with standard HCM methods, which
is truncated by the dimensions of the local stable island.  We identify
one possible halo object likely to become unstable on short timescales,
which could be an example of ongoing dynamical erosion of an asteroid
family.

Finally, we showed that the distribution of
Tina's objects at low albedo ($p_V < 0.12$) in proper $a$ satisfies the null
hypothesis of compatibility with the distribution of objects at high
albedo ($p_V  > 0.12$) when a Kolmogorov-Smirnov test is applied.
This suggests that the two populations are compatible with a common
origin.  The Tina family could be characterized by a rather unique albedo
distribution among X-type asteroid families.

\section*{Acknowledgments}

We thank an anonymous reviewer for valuable comments and suggestions.
This research has made use of the Asteroid Families Portal maintained at
the Department of Astronomy/University of Belgrade.
We would like to thank the Brazilian National Research Council
(CNPq, grant 301577/2017-0), the Foundation for Research Support of
S\~{a}o Paulo state (FAPESP, grant 2016/024561-0),
and the Coordination for the Improvement of
Higher Education Personnel (CAPES, grant 88887.374148/2019-00).
This is a publication from the MASB
(Machine-learning applied to small bodies, 
https://valeriocarruba.github.\-io/Site-MASB/) research group.  Questions on
this paper can also be sent to the group email address:
{\it mlasb2021@gmail.com}.

\section{Code availability}

The codes used for the image classification \citep{2021MNRAS.504..692C}
and identification of family members among the multi-opposition asteroids
\citep{2020MNRAS.496..540C}  are publically available at these GitHub
repositories:

\begin{enumerate}

\item https://github.com/valeriocarruba/ANN$\_$Classification$\_$of$\_\-$M12$\_$resonant$\_$argument$\_$images

\item https://github.com/valeriocarruba/Machine-learning-classification-of-new-asteroid-families-members
  
\section{Data availability}

A database with all the images of asteroids' resonant arguments used
in this work, and their classification, is available at this
repository:  

\vspace{0.15cm}
https://drive.google.com/drive/folders/\-1oxdXquibdYYP965AgzoMjFop-ZZXcHni.

\end{enumerate}

\bibliographystyle{mnras}
\bibliography{mybib}

\bsp

\label{lastpage}

\end{document}